\documentstyle[11pt,amstex,epsf,aps,preprint]{revtex}
\tightenlines

\makeatletter 
\def\teken{\mathop{\operator@font teken}\nolimits}
\def\sign{\mathop{\operator@font sign}\nolimits}
\def\im{\mathop{\operator@font Im}\nolimits}
\def\re{\mathop{\operator@font Re}\nolimits}
\def\Sp{\mathop{\operator@font Sp}\nolimits} 
\def\Tr{\mathop{\operator@font Tr}\nolimits} 
\def\var{\mathop{\operator@font var}\nolimits} 
\def\covar{\mathop{\operator@font covar}\nolimits}

\makeatother

\def\EXPECT#1{\langle #1 \rangle}
\def\KET#1{\vert{#1}\rangle}

\def\magweg#1{}

\def\perm{{\varrho}}

\def\empavg#1{\overline{#1}}
\def\thavg#1{E\left({#1}\right)}

\begin{document}
%\draft
\title{Fast Algorithm for Finding the Eigenvalue Distribution of Very Large Matrices}

\author{Anthony HAMS and Hans De RAEDT}

\address{Institute for Theoretical Physics and
Materials Science Centre,\\
University of Groningen, Nijenborgh 4,
NL-9747 AG Groningen, The Netherlands}

\date{\today}
\maketitle
\begin{abstract}%
A theoretical analysis is given of the equation of motion method, due to Alben et al.,
to compute the eigenvalue distribution (density of states) of very large matrices.
The salient feature of this method is that
for matrices of the kind encountered in quantum physics the memory and CPU
requirements of this method scale linearly with the dimension of the matrix.
We derive a rigorous estimate of the statistical error, supporting
earlier observations that the computational efficiency
of this approach increases with matrix size.
We use this method and an imaginary-time version of it to compute
the energy and the specific heat of three different,
exactly solvable, spin-1/2 models and compare with the exact results
to study the dependence of the statistical errors on sample and matrix size.

\end{abstract}

\narrowtext

\section{Introduction}

The calculation of the distribution of eigenvalues of very large matrices is
a central problem in quantum physics. This distribution determines
the thermodynamic properties of the system (see below).  It is directly
related to the single-particle density of states (DOS) or Green's function.
In a one-particle (e.g., one-electron) description knowledge of the DOS
suffices to compute the transport properties~\cite{Mahan}.

The most direct method to compute the DOS, i.e. all the eigenvalues,
is to diagonalize the matrix $H$ representing the Hamiltonian of the system.
This approach has two obvious limitations: The number of operations increases
as the third power of the dimension $D$ of $H$ and, perhaps most importantly,
the amount of memory required by state-of-the-art algorithms 
grows as $D^2$~\cite{WILKINSON,Golub}. This scaling behavior limits the application
of this approach to matrices of dimension $D={\cal O}(10000)$, which
is too small for many problems of interest. What is needed are methods
that scale linearly with $D$.

There has been considerable interest in developing ``fast'' (i.e. ${\cal O}(D)$)
algorithms to compute the DOS and other similar quantities.
One such algorithm and an application of it
to electron motion in disordered alloy models was given by Alben et al.~\cite{ALBEN}.
In this approach the DOS is obtained by solving
the time-dependent Schr\"odinger equation (TDSE)
of a particle moving on a lattice, followed
by a Fourier transform of the retarded Green's function~\cite{ALBEN}.
Using the unconditionally stable split-step Fast Fourier Transform (FFT)
method to solve the TDSE,
it was shown that the eigenvalue spectrum of a particle moving in continuum
space can be computed in the same manner~\cite{FEIT82}.
Fast algorithms of this kind proved useful to study various
aspects of localization of waves~\cite{HANS89,KAWA96,OHT97} and
other one-particle problems~\cite{IITAKA98,IITAKA99}.
Application of these ideas to quantum many-body systems
triggered further development of flexible and efficient methods
to solve the TDSE. Based on Suzuki's product formula approach,
an unconditionally stable algorithm was developed and used
to compute the time-evolution of two-dimensional
S=1/2 Heisenberg-like models~\cite{PEDROone}.
Results for the DOS of matrices of dimension $D\approx1000000$ where
reported~\cite{PEDROone}.
A potentially interesting feature of these fast algorithms is that
they may run very efficiently on a quantum computer~\cite{hdr2,abramslloyd}.

A common feature of these fast algorithms is that they solve
the TDSE for a sample of randomly chosen initial states.
The efficiency of this approach as a whole relies on the
hypothesis (suggested by the central limit theorem)
that satisfactory accuracy can be achieved by using a small
sample of initial states.
Experience not only shows that this hypothesis is correct, it strongly suggests that
for a fixed sample size the statistical error on physical
quantities such as the energy and specific heat decreases with
the dimension $D$ of the Hilbert space~\cite{hdr2}.

In view of the general applicability of these fast algorithms
to a wide variety of quantum problems it seems warranted to
analyze in detail their properties and the peculiar $D$ dependence in
particular.
In Sections~\ref{sec:theory} and ~\ref{sec:realtime}
we recapitulate the essence of the approach.
We present a rigorous estimate for
the mean square error (variance) on the trace of a matrix.
In Section~\ref{sec:imagtime} we describe
the imaginary-time version of the method.
The statistical analysis of the numerical data is discussed
in Section~\ref{sec:errors}.
Section~\ref{sec:exact} describes
the model systems that are used in our numerical experiments.
The algorithm used to solve the TDSE is reviewed in
Section~\ref{sec:timestep}.
In Section~\ref{sec:bounds} we derive rigorous bounds on the
accuracy with which all eigenvalues can be determined and demonstrate
that this accuracy decreases linearly with the time
over which the TDSE is solved.
The results of our numerical
calculations are presented in Section~\ref{sec:results} and
our conclusions are given in Section~\ref{sec:conclusions}.

\section{Theory} \label{sec:theory}
The trace of a matrix ${A}$ acting on a $D$-dimensional Hilbert-space
spanned by an orthonormal set of states $\{ \KET{\phi_n} \}$ is given by
\begin{align} \label{eq:trdef}
\Tr\, { A} = \sum_{n=1}^{D} \EXPECT{\phi_n \vert A\phi_n}.
\end{align}
Note that according to~(\ref{eq:trdef}) we have $\Tr\,1=D$.
If $D$ is very large one might think
of approximating Eq.~(\ref{eq:trdef}) by sampling over a subset of
$K$ ($K\ll D$)``important'' basis vectors.
The problem with this approach is that the notion ``important''
may be very model-dependent.
Therefore it is better to sample in a different manner.
We construct a random vector $\KET{\psi}$ by choosing $D$
complex random numbers, $c_n\equiv f_n+ig_n$, with mean $0$, for $n=1\ldots D$,
so 
\begin{align} \label{eq:randvec}
\KET{\psi} = \sum_{n=1}^{D} c_n \KET{\phi_n},
\end{align} 
and calculate
\begin{align}
\EXPECT{\psi \vert A\psi}
 = \sum_{n,m=1}^{D} c^*_m c^{\phantom{*}}_n
\EXPECT{\phi_m \vert A\phi_n}.
\end{align}
If we now sample over $S$ realizations of the random vectors $\{\psi\}$
and calculate the average, we obtain
\begin{align}
%\empavg
{1\over S}\sum_{p=1}^{S}
\EXPECT{\psi_p \vert A \psi_p}
=
{1\over S}\sum_{p=1}^{S}\sum_{n,m=1}^{D} %\empavg
{c^*_{m,p} c^{\phantom{*}}_{n,p}}
\EXPECT{\phi_m \vert A \phi_n}.
\end{align}
Assuming that there is no correlation between the
random numbers in different realizations and
that the random numbers $f_{n,p}$ and $g_{n,p}$ are drawn from
an even and symmetric (both with respect to each variable)
probability distribution (see Appendix~\ref{app:expect} for more details), we have
\begin{align} \label{eq:diaglimit}
\lim_{{ S} \rightarrow \infty} %\empavg
{1\over S}\sum_{p=1}^{S} {c^*_{m,p} c^{\phantom{*}}_{n,p}}
= \thavg{|c|^2} \, \delta_{m,n},
\end{align}
where $\thavg{.}$ denotes the expectation value
with respect to the probability distribution used to generate the $c_{n,p}$'s.
In the r.h.s of~(\ref{eq:diaglimit}) the subscripts of $c_{n,p}$ have been dropped
to indicate that the expectation value does not depend on $n$ or $p$.
It follows immediately that
\begin{align}
%\empavg
\lim_{{ S} \rightarrow \infty} %\empavg
{1\over S}\sum_{p=1}^{S}
\EXPECT{\psi_p \vert A \psi_p}
 =
\thavg{|c|^2} \Tr\, { A} = 
\thavg{|c|^2} 
\sum_{n=1}^{D} 
\EXPECT{\phi_n \vert A \phi_n}
,
\end{align}
showing that we can compute the trace of $A$ by sampling
over random states $\{\psi_p\}$, provided there is an efficient
algorithm to calculate 
$\EXPECT{\psi_p \vert A \psi_p}$
%$\BRA{\phi_n} \KET{A\phi_m}$ for any pair $(n,m)$
(see Section~\ref{sec:timestep}).

According to the central limit theorem, for a large but finite ${ S}$ we have
\begin{align}  
\frac{1}{S} \sum_{p=1}^{S} {c^*_{m,p} c^{\phantom{*}}_{n,p}}
= \thavg{|c|^2} \delta_{m,n} + {\cal O} \left(\frac{1}{\sqrt{{ S}}}\right),
\end{align}
meaning  that  the statistical error on the trace
vanishes like $1/\sqrt{{ S}}$, which is not surprising.
What is surprising is that one can prove a much stronger result as follows.
Let us first normalize the $c_{n,p}$'s so that, for all $p$,
\begin{align} \label{eq:annorm}
\sum_{n=1}^D |c_{n,p}|^2 = 1.
\end{align}
This innocent looking step has far reaching consequences.
First we note that the normalization renders
the method exact in (the rather trivial) case that the matrix $A$
is proportional to the unit matrix. The price we pay for this is
that for fixed $p$, the $c_{n,p}$ are now correlated but that
does not cause problems (see Appendix~\ref{app:expect}).
Second it follows that $\thavg{|c|^2} = 1/D$.

Obviously the  error can be written as
\begin{align} \label{eq:rewrite}
 \Tr \, A -  \frac{D}{S} \sum_{p=1}^{S} \EXPECT{\psi_p \vert A \psi_p} 
 = \Tr RA,
\end{align}
where
\begin{align}
R_{m,n} \equiv \delta_{m,n}-
\frac{D}{S} \sum_{p=1}^{S} {c^*_{m,p} c^{\phantom{*}}_{n,p}},
\end{align}
is a traceless (due to Eq.~(\ref{eq:annorm})) Hermitian matrix of random
numbers. We put $X=\Tr R A$ and compute $E(|X|^2)$.
The result for the general case can be found in Appendix~\ref{app:expect}.
For a uniform distribution of the $c_{n,p}$'s on the hyper-sphere defined by
$\sum_{n=1}^D |c_{n,p}|^2 = 1$
the expression simplifies considerably
and we find
\begin{align}
E\left(\left|\Tr RA\right|^2\right)
&= 
\frac{D \Tr A^\dagger A
-  |\Tr A|^2 }{ S (D+1)}  ,
\end{align}
an exact expression for the variance in terms of the sample size $S$, the dimension
$D$ of the matrix $A$ and the (unknown) constants $\Tr A^\dagger A$
and $|\Tr A|$.

Invoking a generalization of Markov's inequality~\cite{GRIMMET}
\begin{align}
{\mathbf P}(|X|^2 \geq a) \leq \frac{E(|X|^2)}{a}
\quad; \quad\forall \, a > 0,
\end{align}
where ${\mathbf P}(Q)$ denotes the probability for the statement $Q$ to be true.
We find that the probability that $|\Tr RA|^2$
exceeds a fraction $a$ of $|\Tr A|^2$ is bounded by
\begin{align}
{\mathbf P}\left(\frac{|\Tr RA|^2}{|\Tr A|^2} \geq a\right)
\leq
\frac{1}{a\, S\, (D+1)}
\frac{D \Tr A^\dagger A  -  |\Tr A|^2 }{|\Tr A|^2}
\quad; \quad\forall \, a > 0,
\end{align}
or, in other words, the relative statistical error $e_A$
on the estimator of the trace of $A$ is given by
\begin{align} \label{eq:eadef}
e_A\equiv
\sqrt{\frac{D \Tr A^\dagger A  -  |\Tr A|^2 }{S (D+1)  |\Tr A|^2} } ,
\end{align}
if $|\Tr A|>0$.
We see that $e_A=0$ if $A$ is proportional to a unit matrix.
From~(\ref{eq:eadef}) it follows that, in general, we may expect $e_A$ to vanish with the
square root of $SD$. The prefactor is a measure for the relative spread
of the eigenvalues of $A$ and is obviously model dependent.
The dependence of $e_A$ on $S$, $D$ and the spectrum of $A$ is
corroborated by the numerical results presented below.

It is also of interest to examine the effect of {\sl not} normalizing the
$c_{n,p}$'s. A calculation similar to the one that lead to the above results
yields
\begin{align}
e_A=
\sqrt{\frac{\Tr A^\dagger A }{S  |\Tr A|^2} } .
\end{align}
Clearly this bound is less sharp and does not vanish if $A$ is
proportional to a unit matrix.

\section{Real-time method} \label{sec:realtime}
The distribution of eigenvalues or density of states (DOS) of
a quantum system is defined as
\begin{equation} \label{eq:dos}
{\cal D}(\epsilon)=\sum_{n=1}^D \delta(\epsilon-E_n) =
\frac{1}{2\pi}\int_{-\infty}^{\infty} e^{it\epsilon}\, \Tr e^{-itH}\, dt
,
\end{equation}
where $H$ is the Hamiltonian of the system and $n$ runs over all the eigenvalues
of $H$. The DOS contains all the 
physical information about the equilibrium properties of the system.
For instance the partition function, the energy, and the heat capacity
are given by
\begin{align} 
Z &= \int_{-\infty}^{\infty} d \epsilon \,{\cal D}(\epsilon)\,e^{-\beta\,\epsilon}, 
\label{eq:z}
\\
E &= \frac{1}{Z} \int_{-\infty}^{\infty} d \epsilon \,\epsilon\,{\cal D}(\epsilon)\,e^{-\beta\,\epsilon}, 
\label{eq:e}
\\
C &= \beta^2 \left(
\frac{1}{Z}\int_{-\infty}^{\infty} d \epsilon \,\epsilon^2\,{\cal D}(\epsilon)\,e^{-\beta\,\epsilon}
-E^2 \right)
\label{eq:c}
, 
\end{align}
respectively.
Here $\beta=1/k_B T$ and $k_B$ is Boltzmann's constant (we put $k_B=1$ and
$\hbar=1$ from now on).

As explained above the trace in the integral~(\ref{eq:dos}) can be
estimated by sampling over random vectors.
For the statistical error analysis discussed below
it is convenient to define a DOS-per-sample by
\begin{equation}
d_p(\epsilon)\equiv
\frac{1}{2\pi}\int_{-\infty}^{\infty} e^{it\epsilon}\, 
\EXPECT{\psi_p \vert e^{-itH} \psi_p}
dt
,
\end{equation}
where the subscript $p$ labels the particular realization of the
random state $\KET{\psi_p}$. The DOS is then given by
\begin{equation}
{\cal D}(\epsilon) = \lim_{S \rightarrow \infty}
\frac{1}{S}\sum_{p=1}^{S} d_p(\epsilon) .
\end{equation}

Schematically the algorithm to compute $d_p(\epsilon)$ consists of the following steps:
\begin{enumerate}
\item Generate a random state $\KET{\psi_p(0)}$, set $t=0$.
\item Copy this state to $\KET{\psi_p(t)}$.
\item Calculate
$\langle \psi_p(0) \vert \psi_p(t) \rangle$ and store the result.
\item Solve the TDSE for a small time step $\tau$, replacing
      $\KET{\psi_p(t)}$ by $\KET{\psi_p(t+\tau)}$
      (see Section~\ref{sec:timestep} for model specific details).
\item Repeat $N$ times from Step 3.
\item Perform a Fourier transform on the tabulated result and store $d_p(\epsilon)$.
\end{enumerate}

In practice the Fourier transform in Eq.~(\ref{eq:dos}) is performed by
the Fast Fourier Transform (FFT).
We use a Gaussian window to account for the finite time $\tau N$ used in the
numerical time-integration of the TDSE. The number of time step
$N$ determines the accuracy
with which the eigenvalues can be computed.
In Section~\ref{sec:bounds}
we prove that this systematic error in the eigenvalues 
vanishes as $1/\tau N$.

Since for any reasonable physical system (or finite matrix)
the smallest eigenvalue $E_0$ is finite, for all practical purposes
$d_p(\epsilon)=0$ for $ \epsilon < \epsilon_0< E_0$.
The value of $\epsilon_0$ is easily determined by
examination of the bottom of spectrum.
To compute $Z$, $E$, or $C$ we simply replace the
interval $[-\infty,+\infty]$ by $[\epsilon_0,+\infty]$.

\section{Imaginary-time Method} \label{sec:imagtime}
The real-time approach has the advantage that it yields
information on all eigenvalues and can be used
to compute both dynamic and static properties without suffering
from numerical instabilities.
However for the computation of the thermodynamic properties,
the imaginary-time version is more efficient.
We will use the imaginary-time method as an independent check on the
results obtained by the real-time algorithm.

Repeating the steps that lead to Eq.~(\ref{eq:z}) we find
\begin{align}
Z &= \Tr \exp(-\beta H) \nonumber \\
&= \lim_{S \rightarrow \infty} \frac{1}{S}\sum_{p=1}^{S} 
\EXPECT{\psi_p \vert \exp(-\beta H) \psi_p},
\end{align}
with similar expressions for $E$ and $C$.

Furthermore we have
\begin{align}
\EXPECT{\psi_p \vert H^n e^{- \beta H} \psi_p}
=
\EXPECT{e^{- \beta H/2} \psi_p \vert H^n e^{- \beta H/2} \psi_p}
, 
\end{align}
assuming $H$ is Hermitian.
Therefore we only need to propagate the random state for
an imaginary time $\beta/2$ instead of $\beta$.
Furthermore we do not need to perform an FFT.
Disregarding these minor differences,
the algorithm is the same as in the real-time case
with $\tau$ replaced by $-i \tau$.

\section{Error Analysis} \label{sec:errors}
Estimating the statistical error on the partition function
$Z$ is easy because it depends linearly on the trace of the (imaginary)
time evolution operator. However the error on $E$ and $C$
depends on this trace in a more complicated manner and this fact
has to be taken into account.

First we define 
\begin{align}
z_p &\equiv \int_{\epsilon_0}^\infty d \epsilon\, d_p(\epsilon)\, e^{-\beta \epsilon}, \\
h_p &\equiv \int_{\epsilon_0}^\infty d \epsilon\, d_p(\epsilon)\, \epsilon \,
e^{-\beta \epsilon}, \\
w_p &\equiv \int_{\epsilon_0}^\infty d \epsilon \,d_p(\epsilon)\, \epsilon^2\, e^{-\beta \epsilon},
\end{align}
for the real-time method and
\begin{align}
z_p &\equiv \EXPECT{\psi_p \vert e^{-\beta H} \psi_p} , \\
h_p &\equiv \EXPECT{\psi_p \vert H e^{-\beta H} \psi_p} , \\
w_p &\equiv \EXPECT{\psi_p \vert H^2 e^{-\beta H} \psi_p} , 
\end{align}
for the imaginary-time method.

For each value of $\beta$ we generate the data $\{z_p\}$, $\{h_p\}$, and $\{w_p\}$,
for $p=1,\ldots,S$. For both cases we have
\begin{align}
Z &= \lim_{S \rightarrow \infty} \empavg{z}, \\
E &= \lim_{S \rightarrow \infty} \frac{\empavg{h}}{\empavg{z}}, \\
C &= \lim_{S \rightarrow \infty} \beta^2 
\left(\frac{\empavg{w}}{\empavg {z}}- \frac{\empavg{h}^2}{\empavg{z}^2}\right),
\end{align}
where $\empavg{x}\equiv S^{-1}\sum_{p=1}^{S} x_p$.
The standard deviations on $\empavg{z}$, $\empavg{h}$, and $\empavg{w}$ are given by
\begin{align}
\delta z &= \sqrt{\frac{\var(z)}{S-1}}, \\
\delta h &= \sqrt{\frac{\var(h)}{S-1}}, \\
\delta w &= \sqrt{\frac{\var(w)}{S-1}}, 
\end{align}
where $\var(x)\equiv\empavg{x^2}-\empavg{x}^2$ denotes the variance on the
data $\{x_p\}$.
However the sets of data $\{z_p\}$, $\{h_p\}$ and $\{w_p\}$
are correlated since they are calculated from the same
set $\{\KET{\psi_p}\}$. These correlations in the data are accounted for by
calculating the covariance matrix $M_{k,l}$ ($k,l=1,\ldots,3$)
the elements of which
are given by $\empavg{x_k x_l}-\empavg{x_k}\,\,\empavg{x_l}$, where 
$\{x_1\}$, $\{x_2\}$, and $\{x_3\}$ are a shorthand for
$\{z_p\}$, $\{h_p\}$, and $\{w_p\}$ respectively.
The estimates for the errors in $Z$, $E$ and $C$ are given by
\begin{align}
\label{eq:dzdef}
\delta Z^2 &= \frac{1}{S-1} {\delta z}^2,\\
\delta E^2 &= \frac{1}{S-1} \sum^3_{k,l=1} M_{k,l} 
\frac{d\empavg{E}}{d\,\overline{x_k}} 
\frac{d\empavg{E}}{d\,\overline{x_l}} ,\\
\delta C^2 &= \frac{1}{S-1} \sum^3_{k,l=1} M_{k,l} 
\frac{d\empavg{C}}{d\,\overline{x_k}} 
\frac{d\empavg{C}}{d\,\overline{x_l}} ,
\end{align}
where $\empavg{E}=\empavg{h}/\empavg{z}$ and
and $\empavg{C}=\beta^2(\empavg{w}/\empavg{z}-\empavg{h}^2/\empavg{z}^2)$.

\section{Exactly Solvable Spin $1/2$ Models} \label{sec:exact}
The most direct way to assess the validity of the approach described
above is to carry out numerical experiments on exactly solvable models.
In this paper we consider three different exactly solvable models,
two spin-1/2 chains and a mean-field spin-1/2 model.
The former have a complicated spectrum, the latter has a highly degenerate eigenvalue
distribution. These spin models differ from those studied elsewhere~\cite{PEDROone,hdr2}
in that they belong to the class of integrable systems.

\subsection{Spin chains}

Open spin chains of $L$ sites described by the Hamiltonian
\begin{align}
 \label{eq:deltamodel}
H=-J \sum_{i=1}^{L-1}
( \sigma_i^x \sigma_{i+1}^x + \Delta \sigma_i^y \sigma_{i+1}^y )
- h \sum_{i=1}^L \sigma_i^z ,
\end{align}
where $\sigma_{i}^x$, $\sigma_{i}^y$, and $\sigma_{i}^z$ denote the
Pauli matrices and $J$, $\Delta$ and $h$ are model parameters, can be solved exactly.
They can be reduced to diagonal form
by means of the Jordan-Wigner transformation~\cite{mattis}.
We have
\begin{align}
H = \sum_{i,j=1}^L \left[
c_i^{+} A_{i,j} c^{\phantom{+}}_j + \frac{1}{2} 
\left( c_i^{+} B_{i,j} c^{+}_j +
c_j^{\phantom{+}} B^\ast_{j,i} c^{\phantom{+}}_i \right)\right] + h L,
\end{align}
where $c_{i}^{+}$ and $c_{i}^{\phantom{{+}}}$ are spin-less fermion operators
and
\begin{align}
A_{i,j} &= -J (1+\Delta) (\delta_{i,j-1}+\delta_{i-1,j}) - 2 h \delta_{i,j}, \\
B_{i,j} &= -J (1-\Delta) (\delta_{i,j-1}-\delta_{i-1,j}),
\end{align}
are $L\times L$ matrices.
By further canonical transformation this Hamiltonian can be written as
\begin{align}
H = \sum_{k = 1}^L \Lambda_k \left(n_k-\frac{1}{2}\right) + \frac{1}{2} \Tr A+ h L,
\end{align}
where $n_k$ is the number operator of state $k$ and the $\Lambda_k$'s
are given by the solution of the eigenvalue equation
\begin{align}
(A-B)(A+B)\phi_k = \Lambda^2_k\,\phi_k.
\end{align}
In the general case this eigenvalue problem of the $L\times L$
Hermitian matrix $(A-B)(A+B)$ is most easily solved numerically.
In the present paper we confine ourselves to two limiting cases:
The XY model ($\Delta=1$) and the Ising model in a transverse field ($\Delta=0$).

\subsection{Mean field model}
The Hamiltonian of the mean-field model reads
\begin{align} \label{eq:mfhamil}
 H = -\frac{J}{L} \sum_{i>j =1}^L \vec \sigma_i \cdot \vec \sigma_j  - h \sum_{i=1}^L \sigma^z_i,
\end{align}
and can be rewritten as
\begin{align}
H = -2 \frac{J}{L} \vec S \cdot \vec S  - 2 h S^z + \frac{3}{2} J,
\end{align}
with 
\begin{align}
\vec S &= \frac{1}{2} \sum_{i=1}^L \vec \sigma_i. 
\end{align}
The single spin-$L/2$ Hamiltonian has eigenvalues
\begin{align}
E_{l,m} &= -2 J l (l+1) /L  - 2 h m + \frac{3}{2} J, 
\end{align}
with degeneracy
\begin{align}
n_{l,m} &=
\frac{2 l +1}{L/2+l+1} \left(\begin{array}{c}L \\ L/2-l \end{array}\right).
\end{align}
This rather trivial model serves as a test for the case of highly
degenerate eigenvalues.

\section{Time Evolution} \label{sec:timestep}
For the approach outlined in Sections~\ref{sec:realtime}
and ~\ref{sec:imagtime} to be of practical use
it is necessary that the matrix elements of the exponential of $H$ can be calculated
efficiently. The purpose of this section is to describe how this can be done.

The general form of the Hamiltonians of the models we study is
\begin{equation}
 \label{hamiltonian}
H=-\sum_{i,j=1}^L\sum_{\alpha=x,y,z} J_{i,j}^\alpha \sigma_i^\alpha \sigma_j^\alpha
-\sum_{i=1}^L\sum_{\alpha=x,y,z} h_{i}^\alpha \sigma_i^\alpha
,
\end{equation}
where the first sum runs over all pairs $P$ of spins,
$\sigma_i^\alpha$ ($\alpha=x,y,z$) denotes the $\alpha$-th component of the spin-1/2
operator representing the $i$-th spin.
For both methods, we have to calculate the evolution of a random state,
i.e. $U(\tau)\KET{\psi} \equiv \exp(-i\tau H) \KET{\psi}$
or $U(\tau)\KET{\psi} \equiv \exp(-\tau H) \KET{\psi}$
for the real and imaginary time method  respectively.
We will discuss the real-time case only, the imaginary-time problem can
be solved in the same manner.

Using the semi-group property $U(t_1)U(t_2)=U(t_1+t_2)$
we can write $U(t)=U(\tau)^m$ where $t=m\tau$. Then the main step is to
replace $U(\tau)$ by a symmetrized product-formula approximation~\cite{HDRCPR}.
For the case at hand it is expedient to take
\begin{align}
U(\tau)\approx {\widetilde U(\tau)} \equiv&
e^{-i\tau H_z/2}
e^{-i\tau H_y/2}
e^{-i\tau H_x}
%\nonumber \\ &\times
e^{-i\tau H_y/2}
e^{-i\tau H_z/2}
,
\end{align}
where
\begin{equation}
H_\alpha=-\sum_{i,j=1}^L J_{i,j}^\alpha \sigma_i^\alpha \sigma_j^\alpha
-\sum_{i=1}^L h_{i}^\alpha \sigma_i^\alpha
\quad;\quad\alpha=x,y,z
.
\end{equation}
Other decompositions~\cite{PEDROone,SUZUKItwo}
work equally well but are somewhat less efficient for the cases at hand.
In the real-time approach ${\widetilde U(\tau)}$ is unitary
and hence the method is unconditionally stable~\cite{HDRCPR}
(also the imaginary-time method can be made unconditionally stable).
It can be shown that
$\Vert U(\tau) - {\widetilde U(\tau)}\Vert \le s \tau^3$
($s>0$ a constant)~\cite{normofamatrix}, implying
that the algorithm is correct to second order in the time step $\tau$~\cite{HDRCPR}.
Usually it is not difficult to choose $\tau$ so small that
for all practical purposes the results obtained can be considered
as being ``exact''.
Moreover, if necessary, ${\widetilde U(\tau)}$
can be used as a building block to construct higher-order 
algorithms~\cite{HDRone,SUZUKItwo2,HDRKRMone,SUZUKIthree}.
In Appendix~\ref{app:bounds} we will derive bounds on the error in the eigenvalues
when they are calculated using a symmetric product formula.

As basis states $\{\KET{\phi_n}\}$ we take
the direct product of the eigenvectors of the
$S_i^z$ (i.e. spin-up $\KET{\uparrow_i}$ and spin-down $\KET{\downarrow_i}$).
In this basis, $e^{-i\tau H_z/2}$  changes the input state by altering
the phase of each of the basis vectors.
As $H_z$ is a sum of pair interactions it is trivial to rewrite this operation
as a direct product of 4x4 diagonal matrices (containing
the interaction-controlled phase shifts) and 4x4 unit matrices.
Still working in the same representation, the action of $e^{-i\tau H_y/2}$ 
can be written in a similar manner but the matrices that contain the
interaction-controlled phase-shift have to be replaced by
non-diagonal matrices. Although this does not present a real problem it is
more efficient and systematic to proceed as follows.
Let us denote by $X$($Y$) the rotation by $\pi/2$ of each spin
about the $x$($y$)-axis. As
\begin{equation}
 \label{last}
e^{-i\tau H_y/2}=XX^\dagger e^{-i\tau H_y/2}XX^\dagger
=X e^{-i\tau H_z^\prime/2}X^\dagger ,
\end{equation}
it is clear that the action of $e^{-i\tau H_y/2}$ can be computed by
applying to each spin, the inverse of $X$
followed by an interaction-controlled phase-shift and $X$ itself.
The prime in~(\ref{last}) indicates that $J_{i,j}^z$ and $h_{i}^z$ in $H_z$
have to be replaced by $J_{i,j}^y$ and $h_{i}^y$ respectively.
A similar procedure is used to compute the action of
$e^{-i\tau H_x}$. We only have to replace $X$ by $Y$.

\section{Accuracy of the computed eigenvalues} \label{sec:bounds}

First we consider the problem of how to choose the number
of time steps $N$ to obtain the DOS with acceptable accuracy.
According to the Nyquist sampling theorem employing a sampling
interval $\Delta t = \pi/\max_i |E_i|$ is sufficient to cover the full range
of eigenvalues.
On the other hand the time step also determines the accuracy
of the approximation ${\widetilde U(\tau)}$. Let us call the maximum
value of $\tau$
which gives satisfactory accuracy $\tau_0$ (for the imaginary-time method,
this is the only parameter).
For the examples treated here $\tau_0< \Delta t)$, implying that
we have to use more steps to solve the TDSE than we actually use
to compute the FFT. Eigenvalues that differ less than
$\Delta \epsilon=\pi/N\Delta t$ cannot be identified properly.
However since $\Delta \epsilon\propto N^{-1}$ we only have to extend
the length of the calculation by a factor of two to increase
the resolution by the same factor.

At first glance the above reasoning may seem to be a little optimistic.
It apparently overlooks the fact that if we integrate the TDSE
over longer and longer times the error on the wave function also increases
(although it remains bounded because of the unconditional
stability of the product formula algorithm).
In fact it has been shown that in general~\cite{HDRCPR}

\begin{equation}
\Vert e^{-itH}\KET{\psi(0)} - {\widetilde U}^m(\tau)\KET{\psi(0)}
\Vert \le c \tau^2 t,
\end{equation}
where $t=m\tau$, suggesting that the loss in accuracy on the wave function
may well compensate for the gain in resolution that we get
by using more data in the Fourier transform.
Fortunately this argument does not apply
when we want to determine the eigenvalues as we now show.
As before we will discuss the real-time algorithm only because
the same reasoning (but different mathematical proofs) holds
for the imaginary-time case.

Consider the time-step operator (52). Using the fact that
any unitary matrix can be written as the matrix exponential of a Hermitian matrix
we can write

\begin{equation}
{\widetilde U(\tau)}=
e^{-i\tau H_z/2}
e^{-i\tau H_y/2}
e^{-i\tau H_x}
e^{-i\tau H_y/2}
e^{-i\tau H_z/2}
\equiv
e^{-i\tau \widetilde H(\tau)}.
\end{equation}
It is clear that in practice the real-time method yields
the spectrum of ${\widetilde H(\tau)}$, not
the one of $H$. Therefore the relevant question is:
How much do the spectra of ${\widetilde H(\tau)}$ and
$H$ differ?
In Appendix B we give a rigorous proof that the
difference between the eigenvalues of
${\widetilde H(\tau)}$ and $H$ vanishes as $\tau^2$.
In other words the value of $m$ (or $t=m\tau$) has
no effect whatsoever on the accuracy with which the spectrum
can be determined.
Therefore the final conclusion is that the error in the eigenvalues
vanishes as $\tau^2/N$ where $N$ is the number of data points used in the
Fourier transform of $\Tr e^{-it \widetilde H(\tau)}$.

\section{Results} \label{sec:results}
In all our calculations we take $J = 1$ and $h=0$,
except for the Ising model in a transverse field, 
where we take $h=0.75$.
The random numbers $c_{n,p}$ are generated such that the conditions
Eqs.~(\ref{eq:cnreq1}) and~(\ref{eq:cnreq2}) are satisfied.
We use two different techniques to generate these random numbers:
\begin{enumerate}
\item A uniform random number generator produces $\{f_{n,p}\}$ and $\{g_{n,p}\}$ with
$-1\leq f_{n,p}, g_{n,p} \leq 1$. We then normalize the vector (see Eq.~(\ref{eq:annorm})).
\item The $c_{n,p}$'s are obtained from a two-variable (real and imaginary part)
Gaussian random number generator and the resulting vector is normalized.
\end{enumerate}
Both methods satisfy the basic requirements Eqs.~(\ref{eq:cnreq1}) and~(\ref{eq:cnreq2})
but because the first samples points out of a $2D$-dimensional hypercube
and subsequently projects the vector onto a sphere, the points are not distributed
uniformly over the surface of the unit hyper-sphere.
The second method is known to generate numbers which are distributed uniformly over
the hyper-surface.
Although the first method does not satisfy all the mathematical conditions
that lead to the error~(\ref{eq:eadef}), our numerical experiments with both generators
give identical results, within statistical errors of course.
Also, within the statistical errors,
the results from the imaginary and real-time algorithm are the same.
Therefore we only show some representative results as obtained
from the real-time algorithm.

\begin{figure}[p]
\begin{center}
\begin{tabular}{ccc}
\epsfxsize=5.3cm
\epsffile{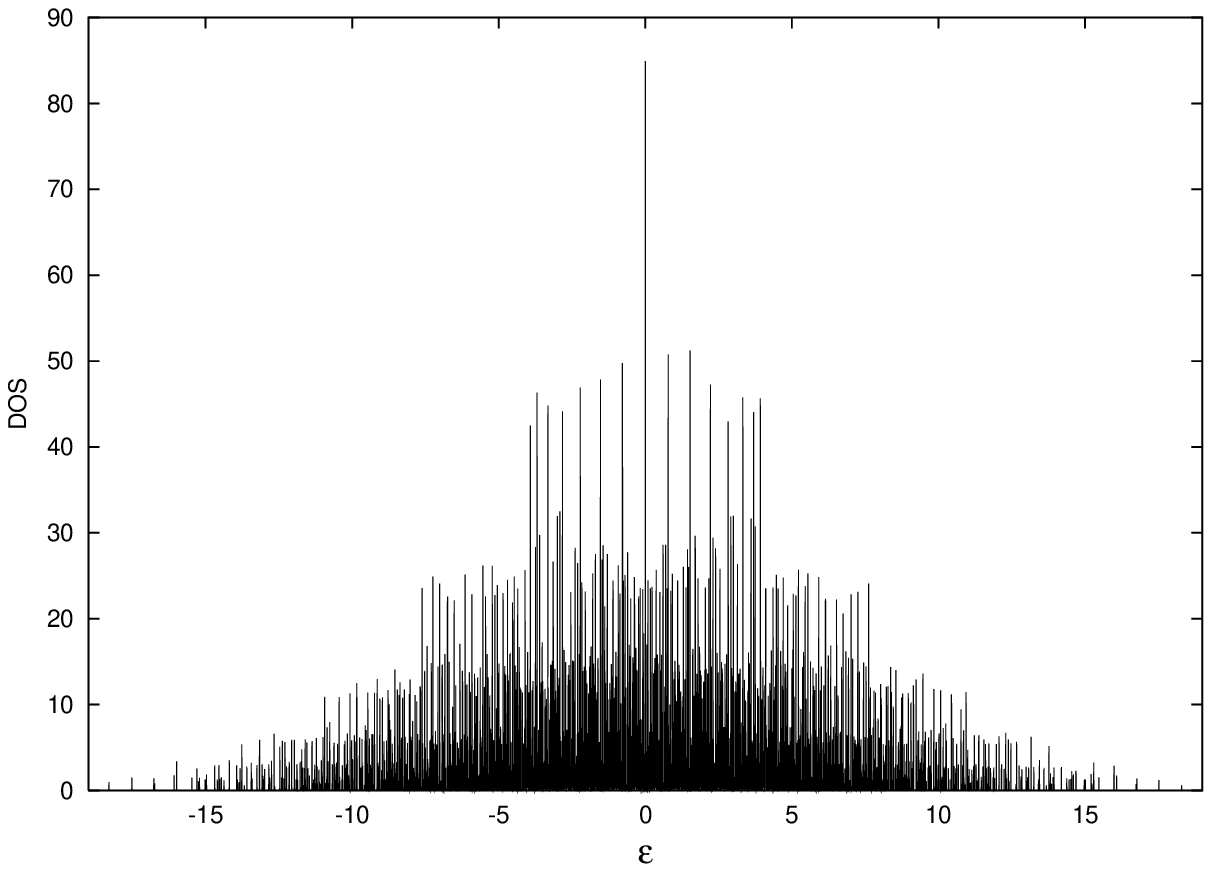}
&
\epsfxsize=5.3cm
\epsffile{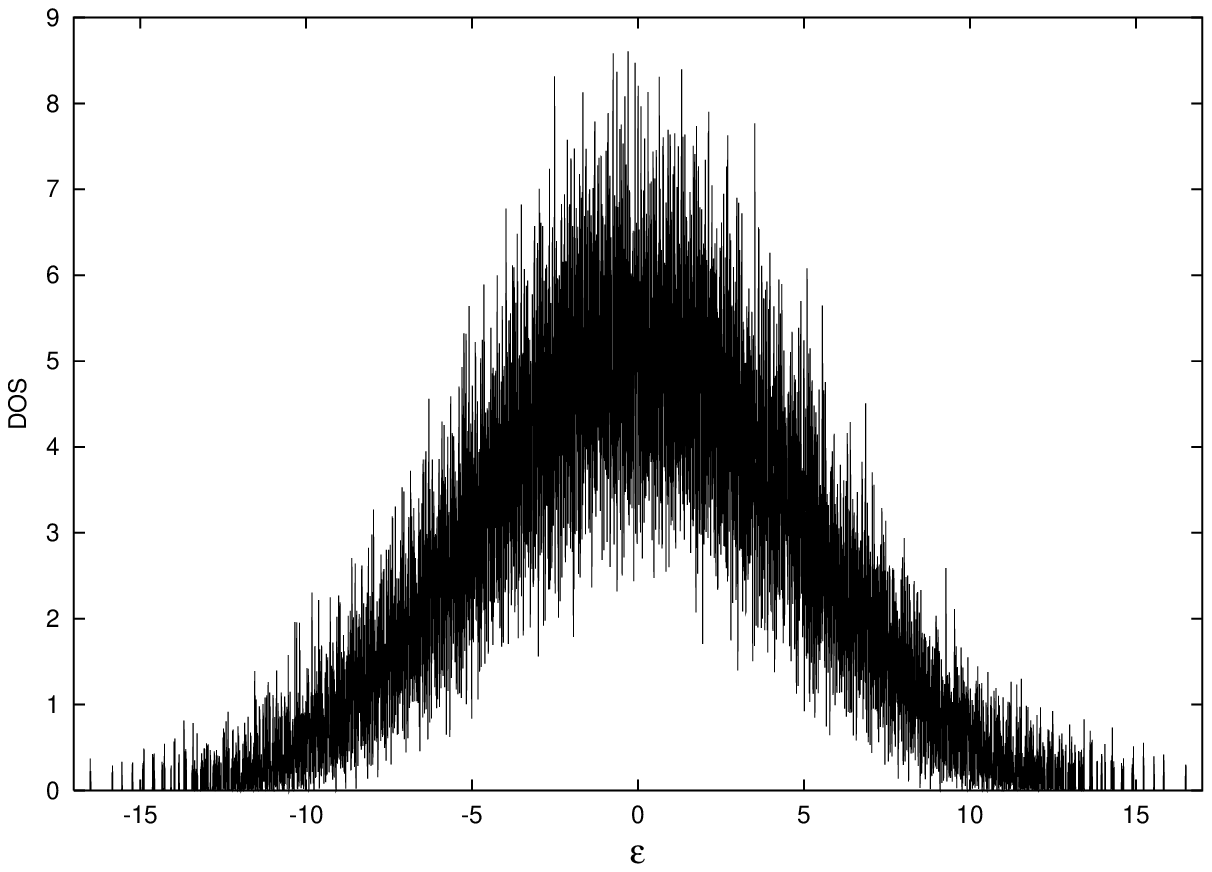}
&
\epsfxsize=5.3cm
\epsffile{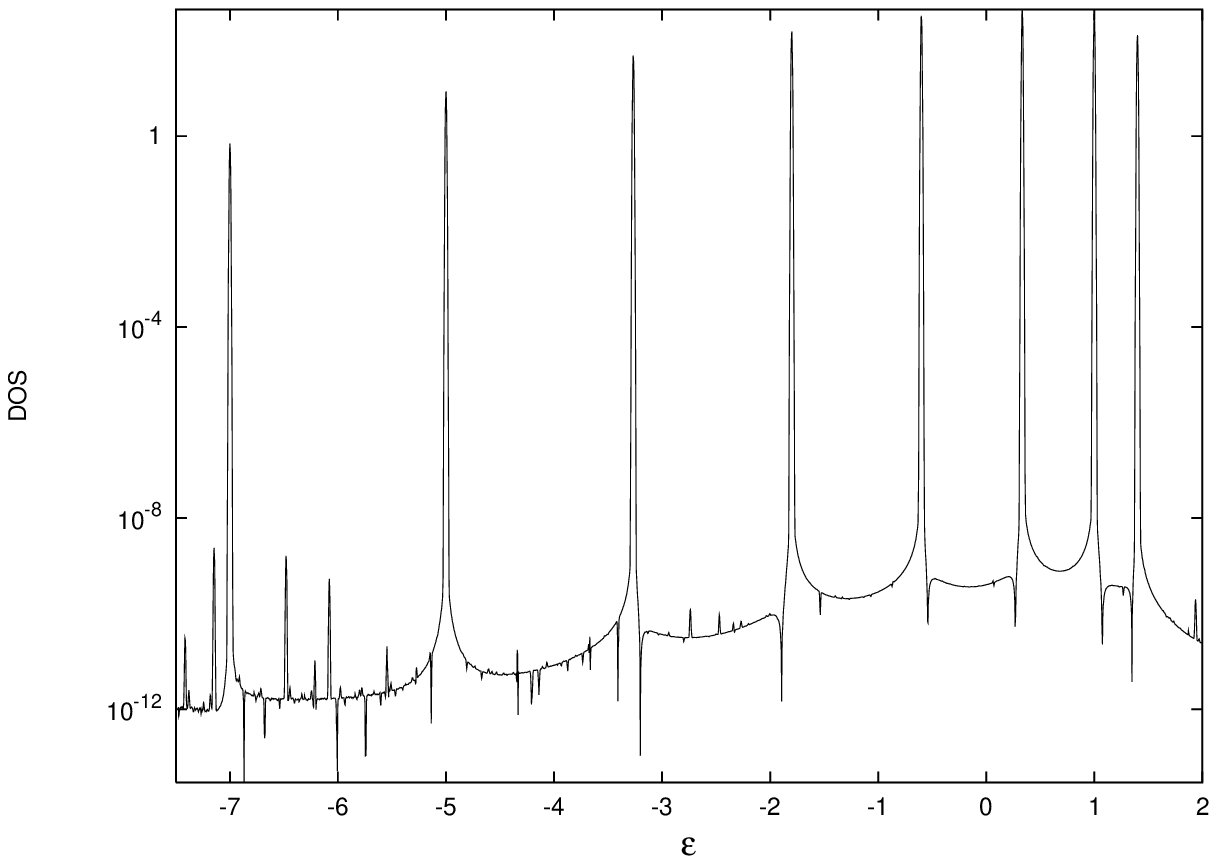}
\end{tabular}
\end{center}
\caption{\noindent%
The density of states (DOS) as obtained from
the real-time algorithm for spin chains of length
$L=15$ and for $S=20$ random initial states.
Left: XY model; middle: Ising model in a transverse field;
right: Mean-field model. For the mean-field model
a logarithmic scale was used to show the highly-degenerate
spectrum more clearly.
}
\label{fig:dos}
\end{figure}

In Fig.~\ref{fig:dos} we show a typical result for the DOS $D(\epsilon)$
of the XY model, the Ising model in a transverse field and the mean-field model,
all with $L=15$ spins and using $S=20$ samples. Because of the very high degeneracy we plotted
the DOS for the mean-field model on a logarithmic scale.

\begin{figure}[p]
\begin{center}
\begin{tabular}{ccc}
\epsfxsize=5.3cm
\epsffile{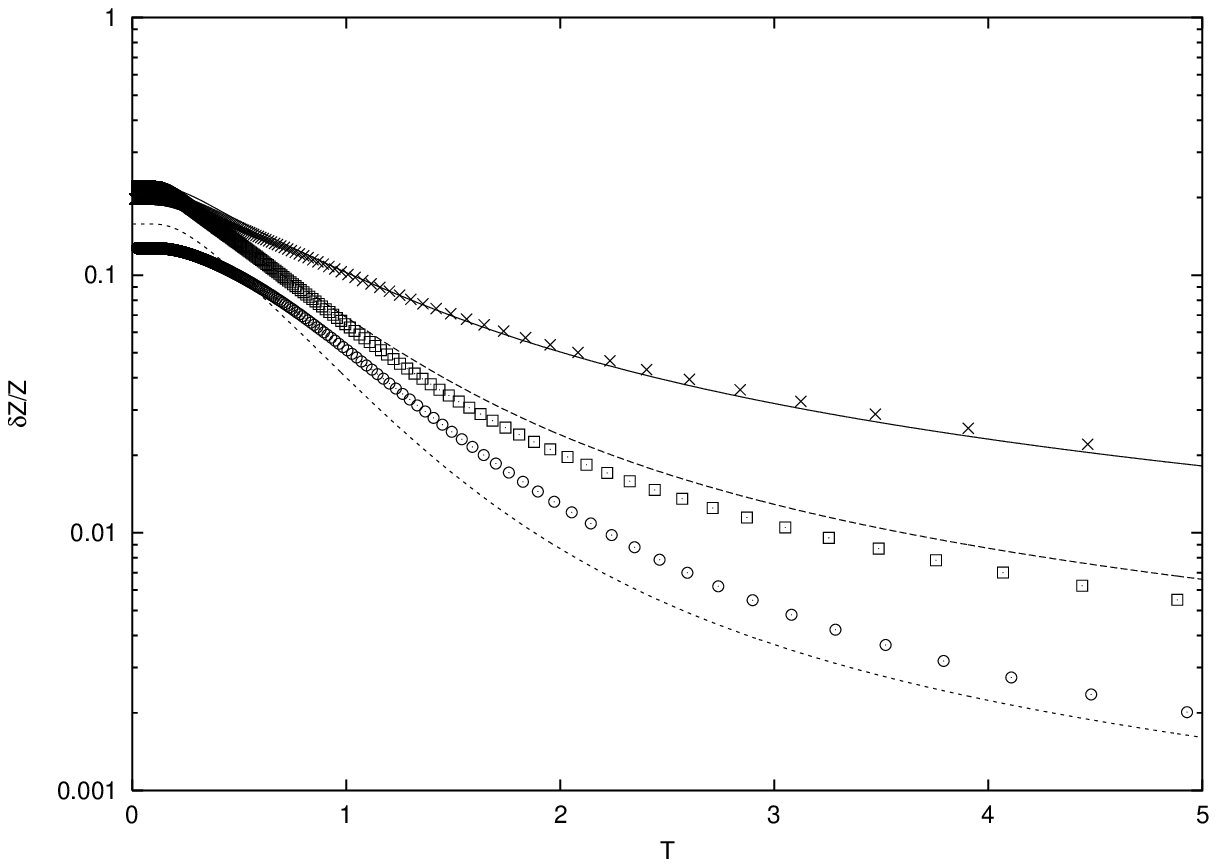}
&
\epsfxsize=5.3cm
\epsffile{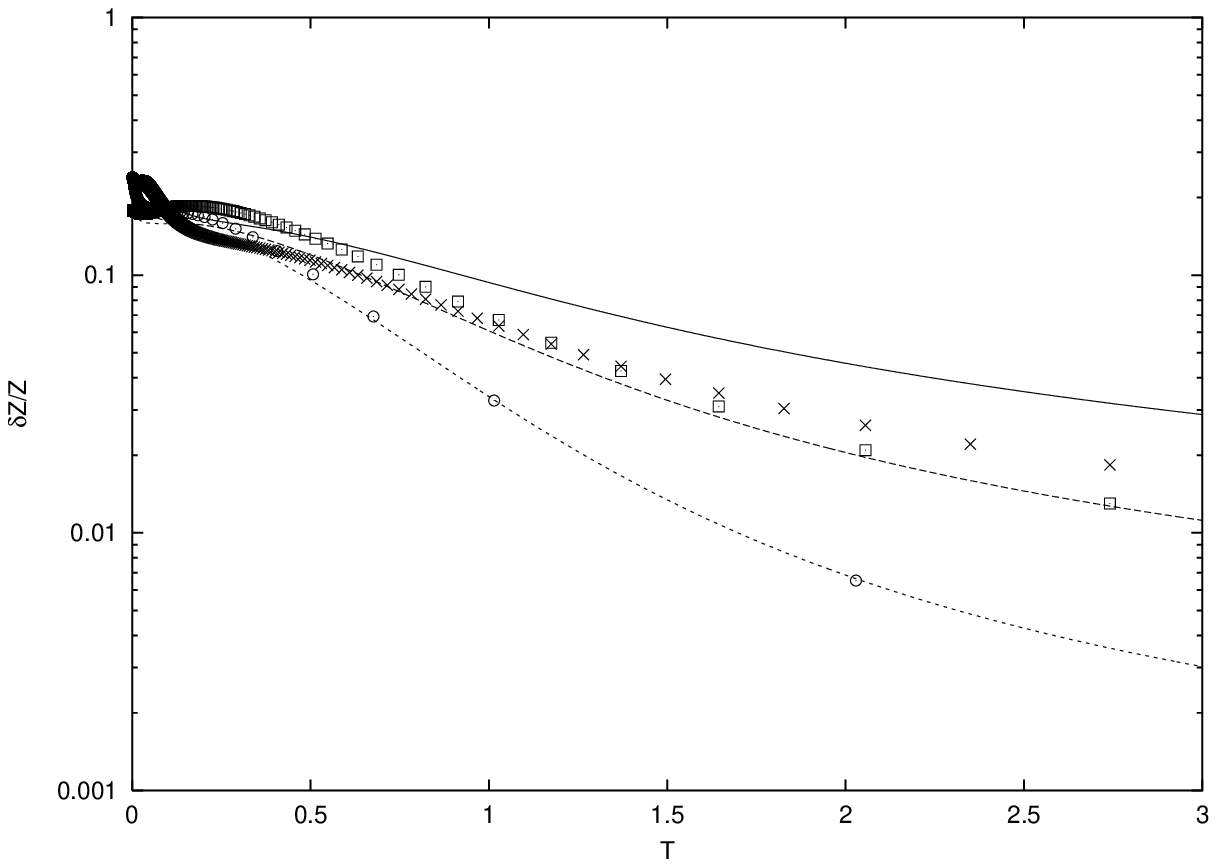}
&
\epsfxsize=5.3cm
\epsffile{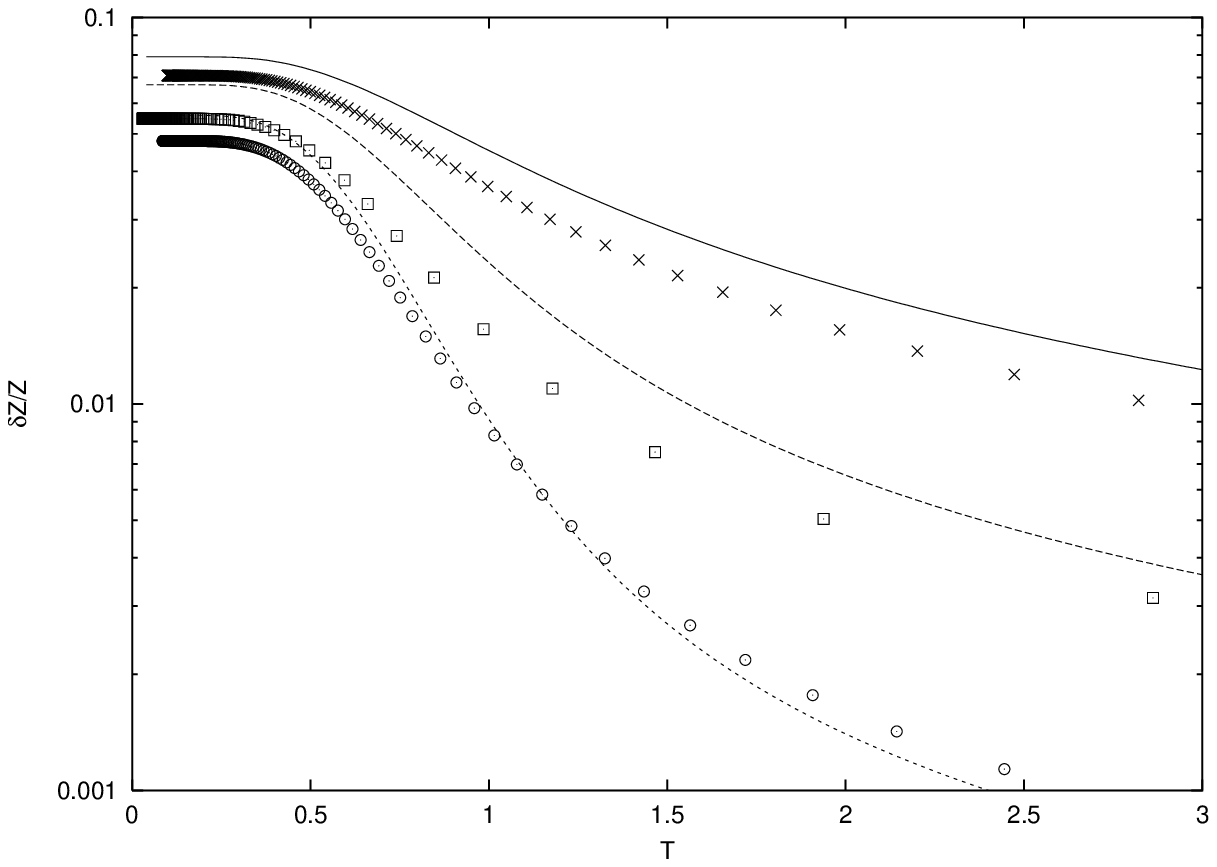}
\end{tabular}
\end{center}
\caption{\noindent%
The relative error $\delta Z/Z$ (see Eq.~(\ref{eq:dzdef})) on a logarithmic scale
as a function of temperature $T\equiv1/\beta$ and for various system sizes.
Left: XY model; middle: Ising model in a transverse field;
right: Mean-field model.
Solid lines: $e_A$ (with $A=e^{-\beta H}$, see Eq.~(\ref{eq:eadef})) for $L=6$;
dashed lines: $e_A$ for $L=10$;
dash-dotted line: $e_A$ for $L=15$.
Crosses: Simulation data for $S=20$ and $L=6$;
squares: Simulation data for $S=20$ and $L=10$;
circles: Simulation data for $S=20$ and $L=15$.
}
\label{fig:dzreal}
\end{figure}

In Fig.~\ref{fig:dzreal}
we show the relative error $\delta Z/Z$ based on Eq.~(\ref{eq:dzdef}) for the three models
of various size, as obtained from the simulation (symbols). For these figures
we used the imaginary-time algorithm, because then the statistical error can be related 
 to $e_A$ directly (see Eq.~(\ref{eq:eadef}) with $A=\exp(-\beta H)$).
The theoretical results (lines) for the error estimate,
obtained by a direct exact numerical evaluation of~(\ref{eq:eadef}) are shown too.
We conclude that for all systems, lattice
sizes and temperatures there is very good agreement
between numerical experiment and theory.

\begin{figure}[p]
\begin{center}
\begin{tabular}{ccc}
\epsfxsize=5.3cm
\epsffile{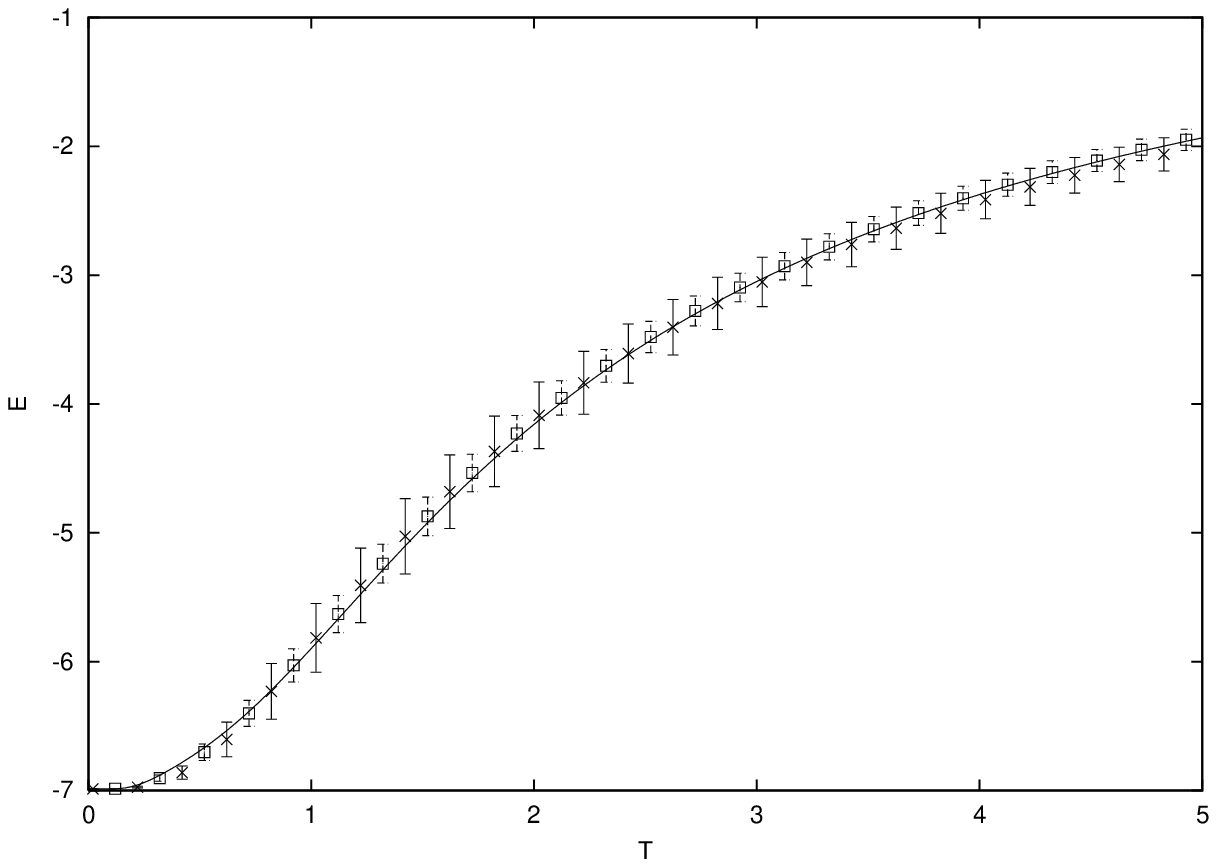}
&
\epsfxsize=5.3cm
\epsffile{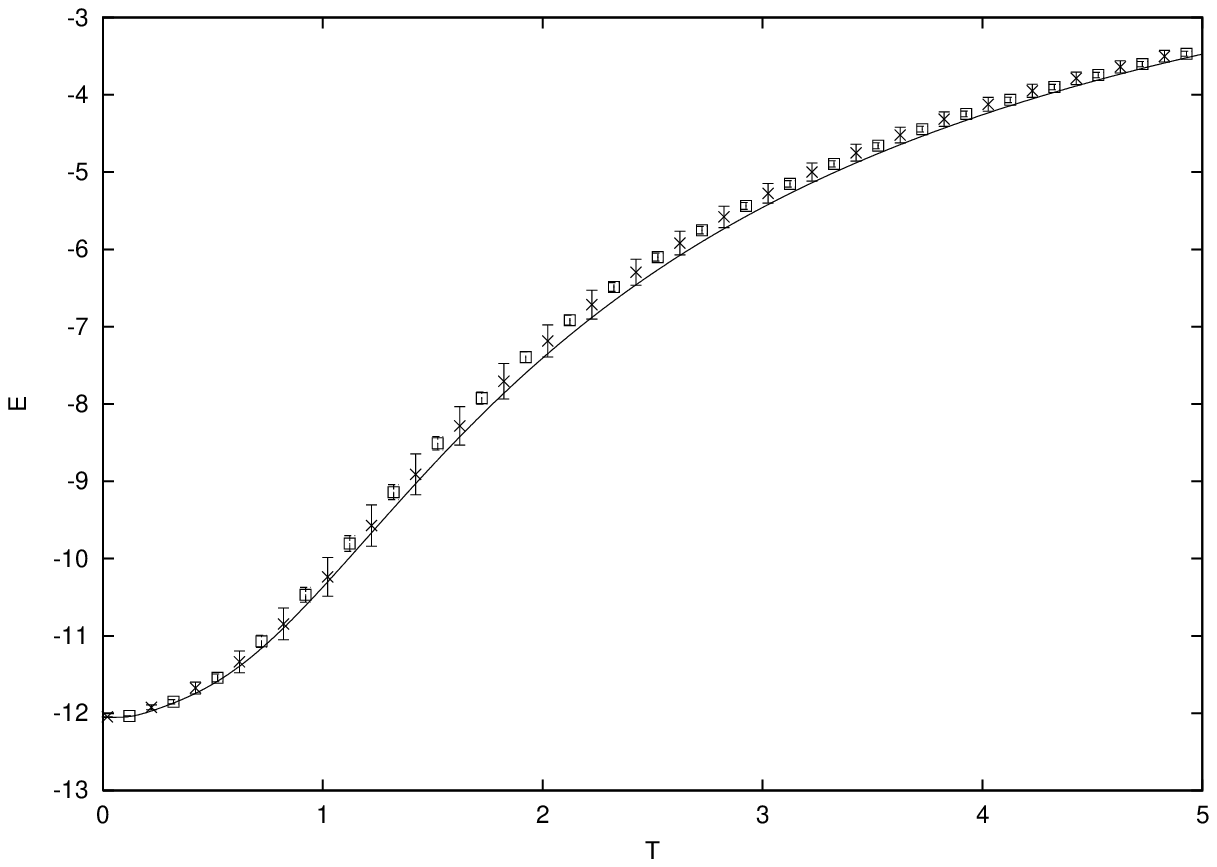}
&
\epsfxsize=5.3cm
\epsffile{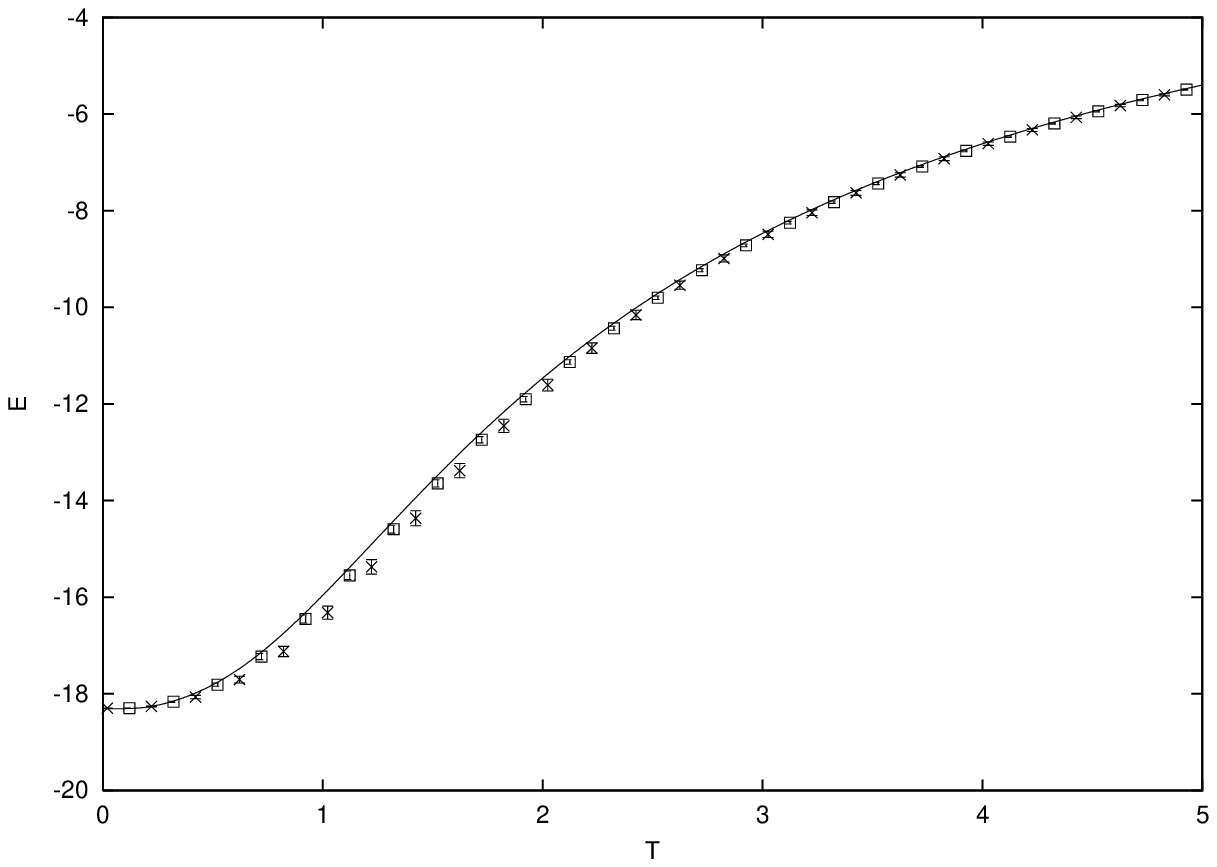}
\\
\epsfxsize=5.3cm
\epsffile{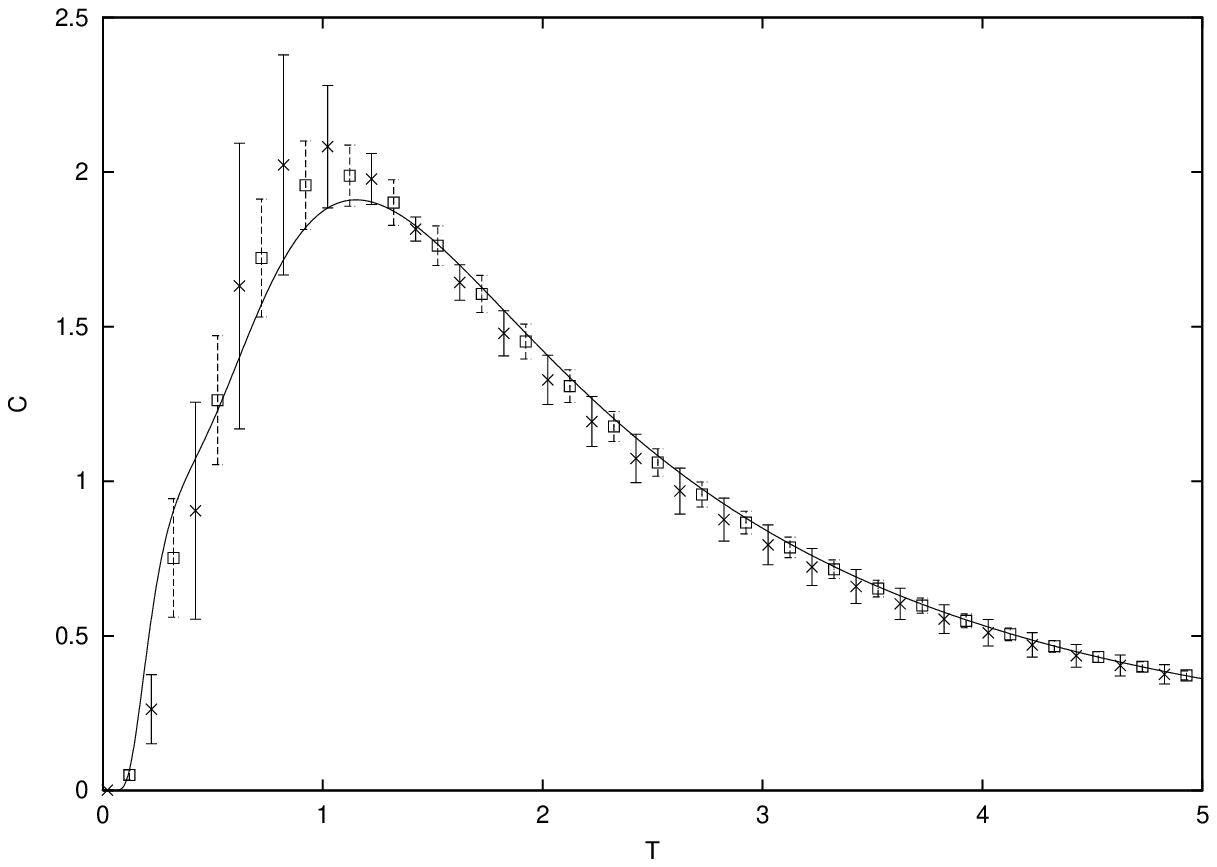}
&
\epsfxsize=5.3cm
\epsffile{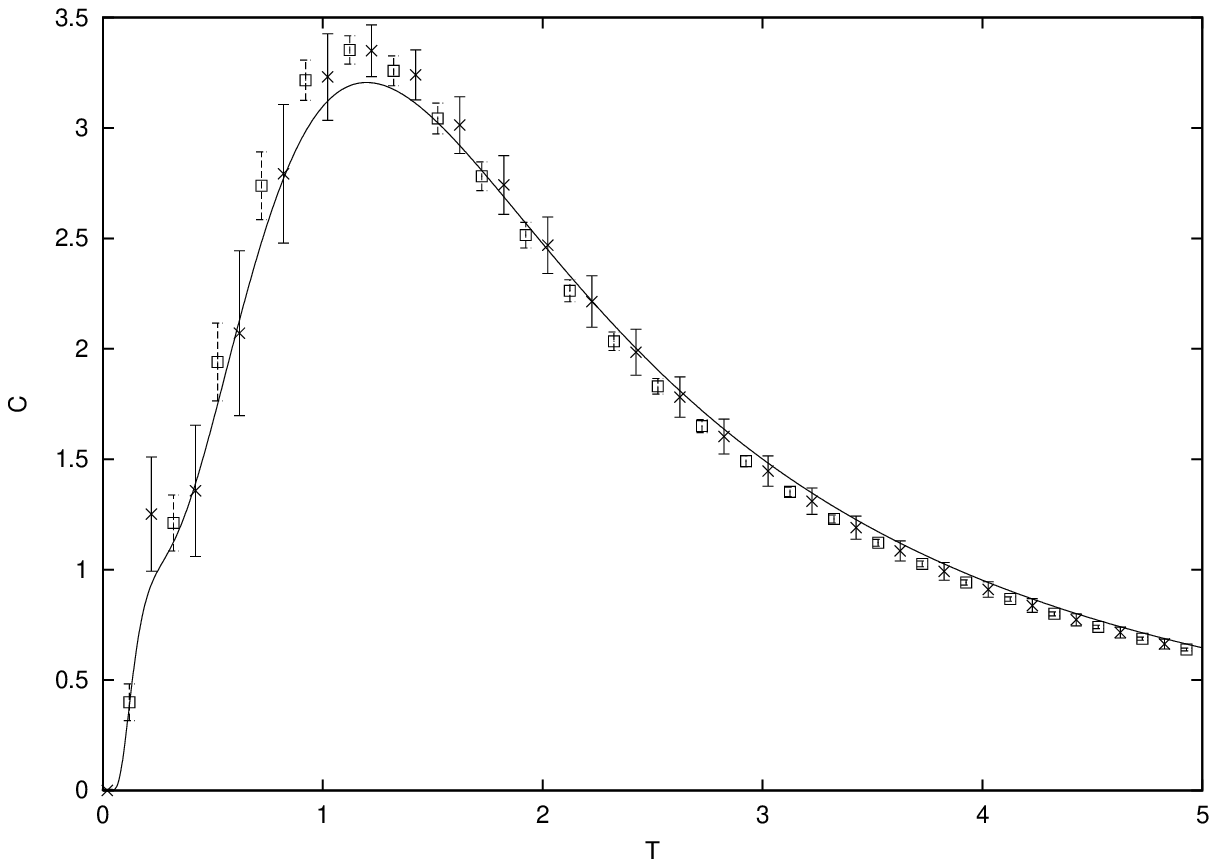}
&
\epsfxsize=5.3cm
\epsffile{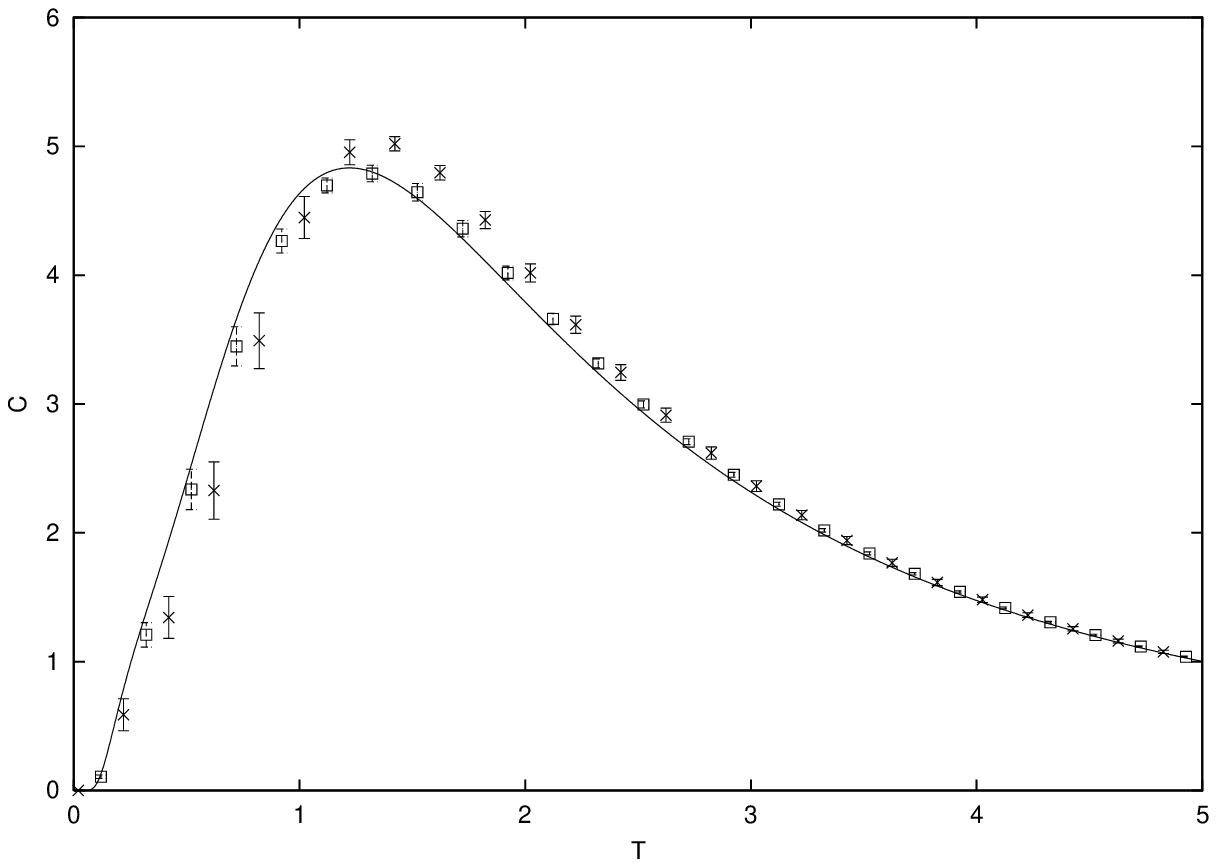}
\end{tabular}
\end{center}
\caption{
Energy (top) and specific heat (bottom) of the XY-model (see~(\ref{eq:deltamodel})) with
$\Delta = 1$, $h = 0$ and $J=1$.
Left: $L=6$; middle: $L=10$; right: $L=15$.
Solid lines: Exact result.
crosses: Simulation data using $S=5$ samples;
squares: Simulation data using $S=20$ samples.
Error bars: One standard deviation.
}
\label{fig:xyreal}
\end{figure}

\begin{figure}[p]
\begin{center}
\begin{tabular}{ccc}
\epsfxsize=5.3cm
\epsffile{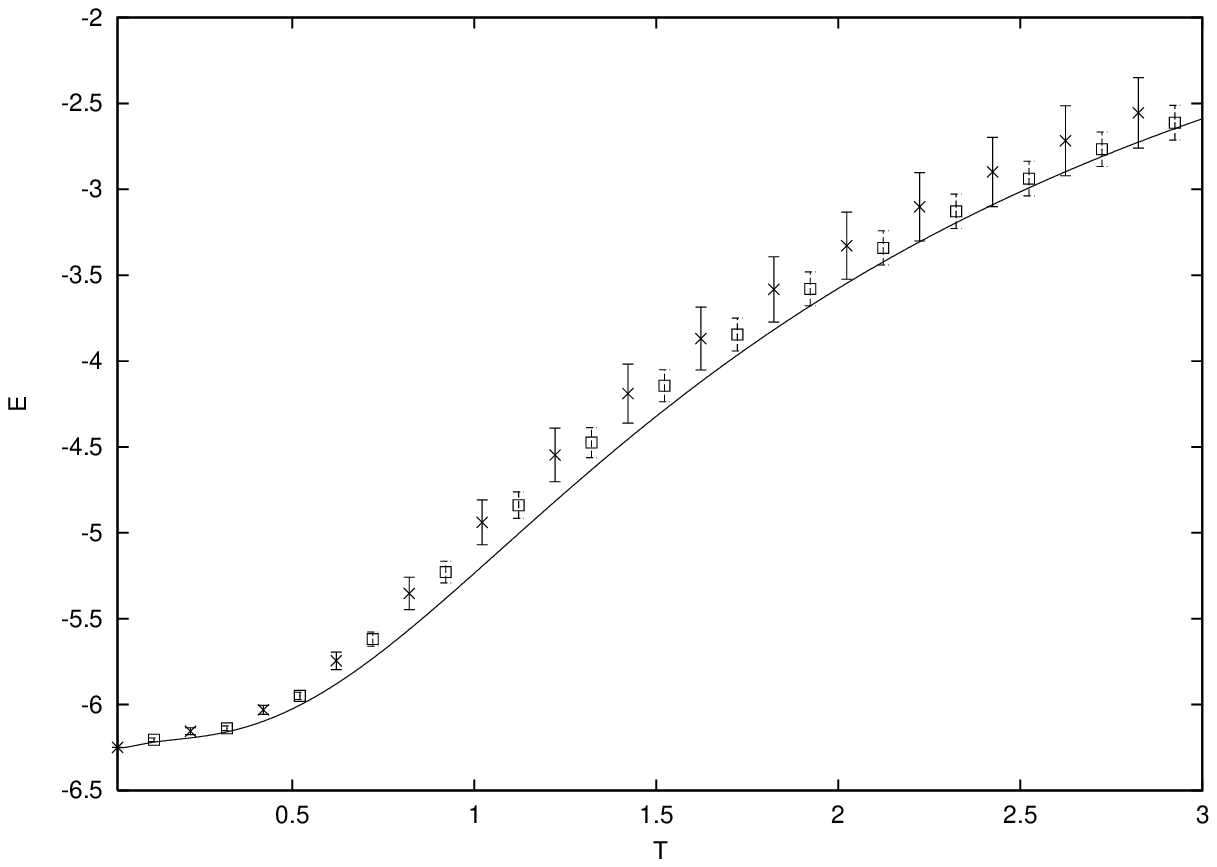}
&
\epsfxsize=5.3cm
\epsffile{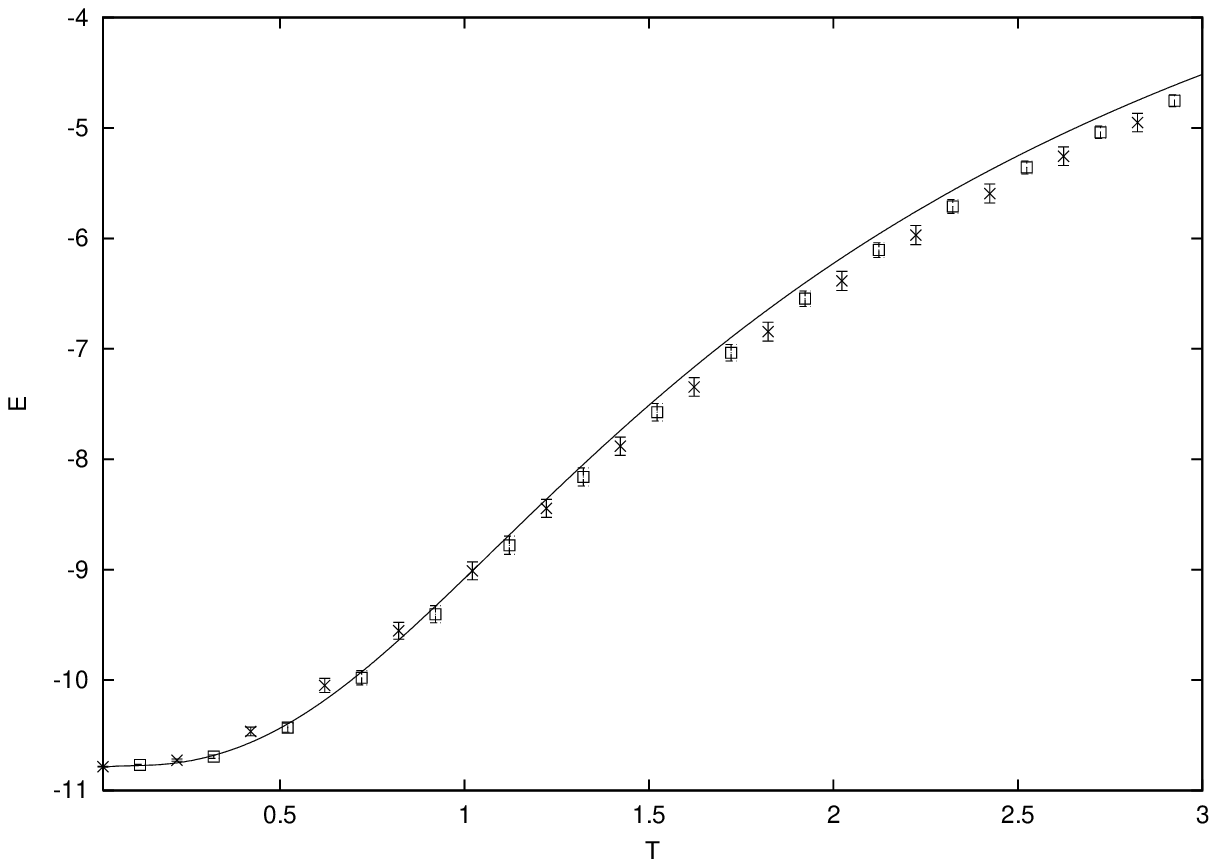}
&
\epsfxsize=5.3cm
\epsffile{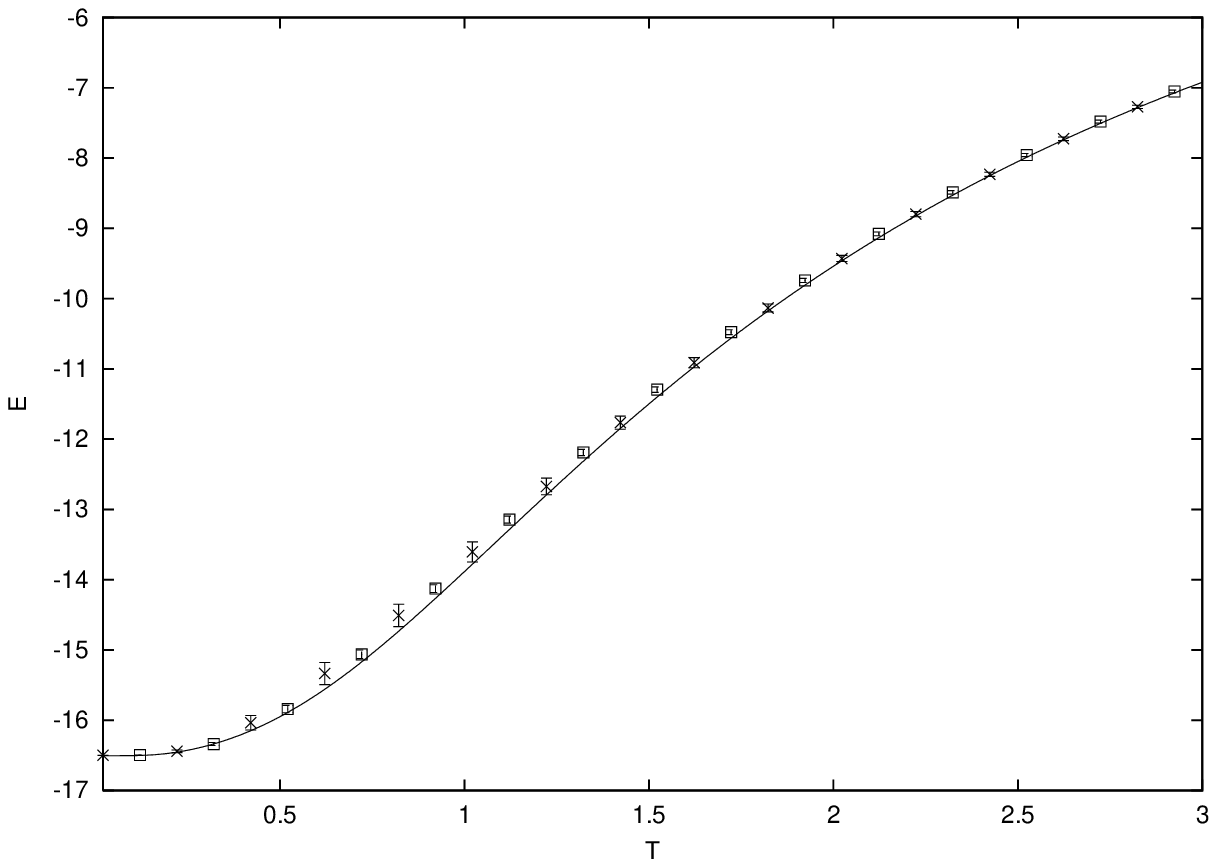}
\\
\epsfxsize=5.3cm
\epsffile{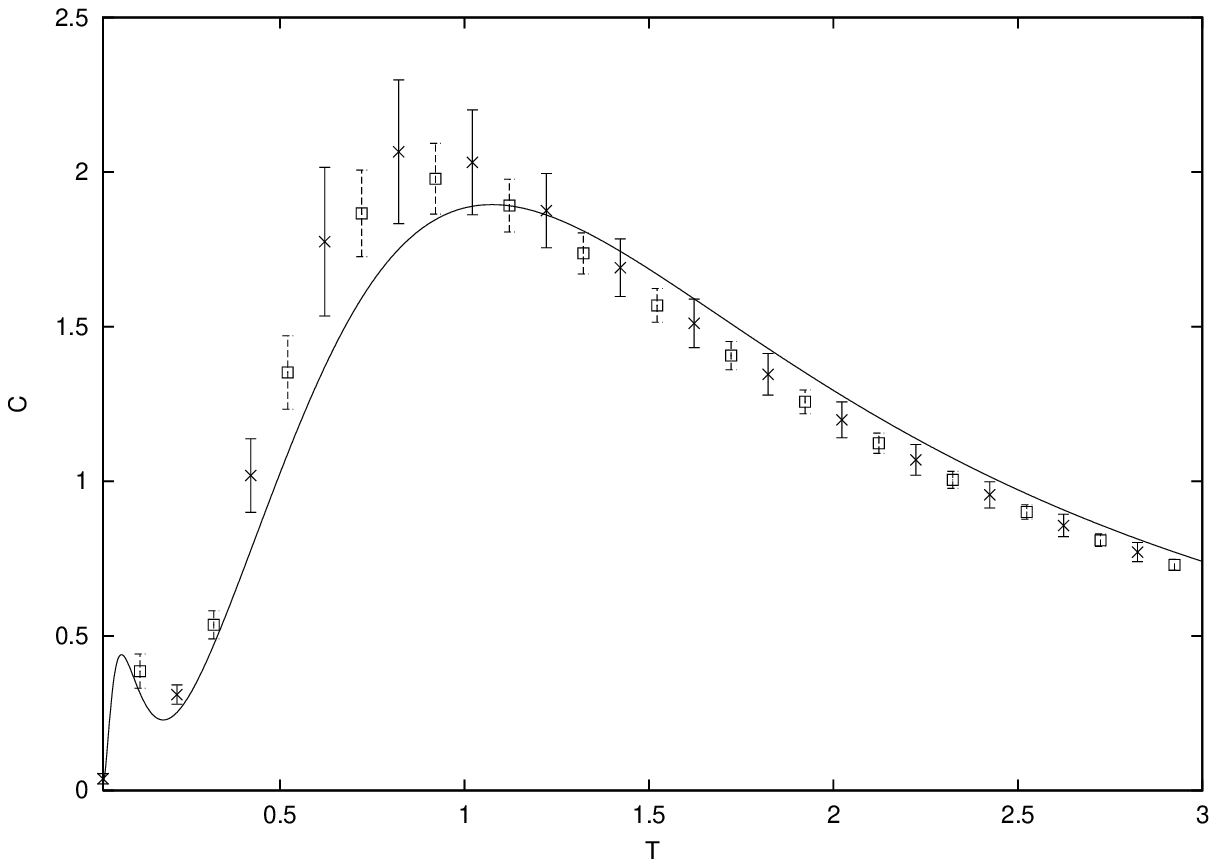}
&
\epsfxsize=5.3cm
\epsffile{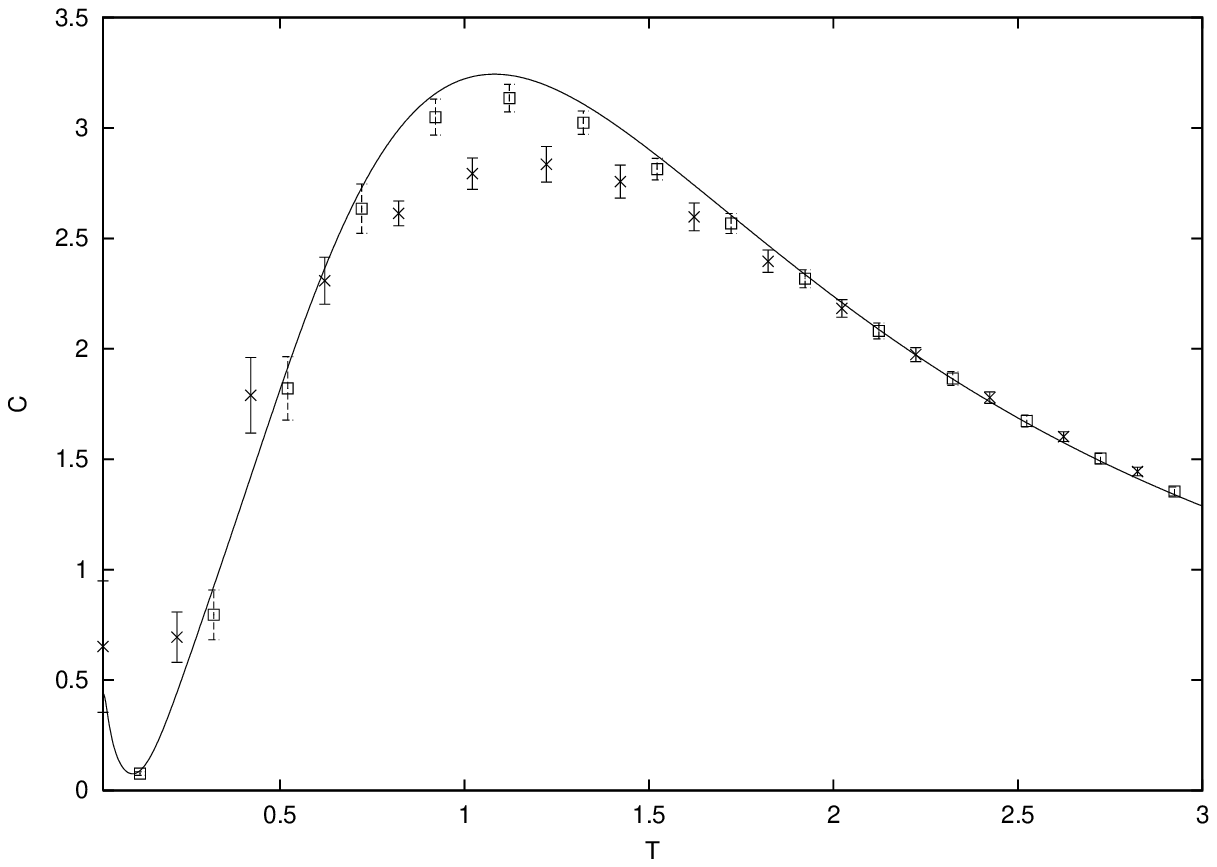}
&
\epsfxsize=5.3cm
\epsffile{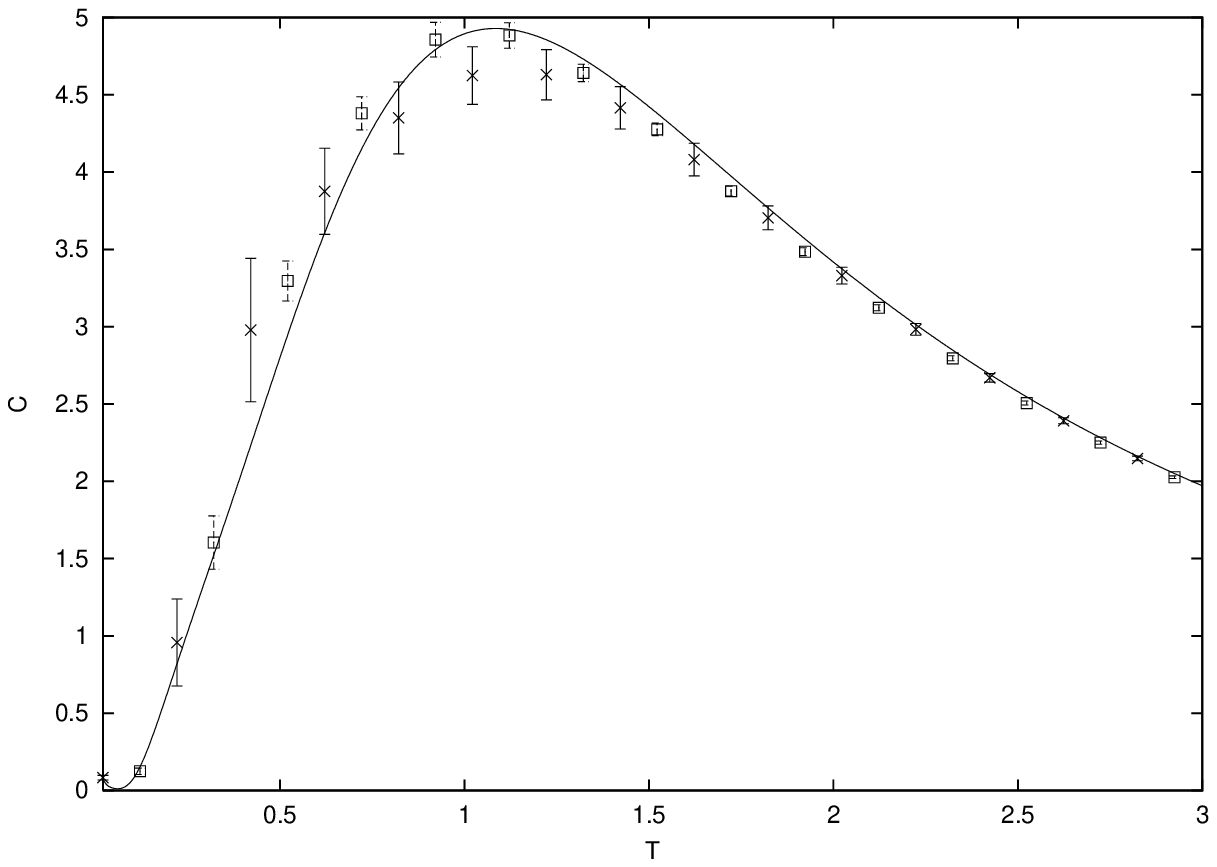}
\end{tabular}
\end{center}
\caption{
Energy (top) and specific heat (bottom)
of the Ising model in a transverse field (see~(\ref{eq:deltamodel}))
with $\Delta = 0$, $h = 0.75$ and $J=1$.
Left: $L=6$; middle: $L=10$; right: $L=15$.
Solid lines: Exact result.
crosses: Simulation data using $S=5$ samples;
squares: Simulation data using $S=20$ samples.
Error bars: One standard deviation.
}
\label{fig:isreal}
\end{figure}

\begin{figure}[p]
\begin{center}
\begin{tabular}{ccc}
\epsfxsize=5.3cm
\epsffile{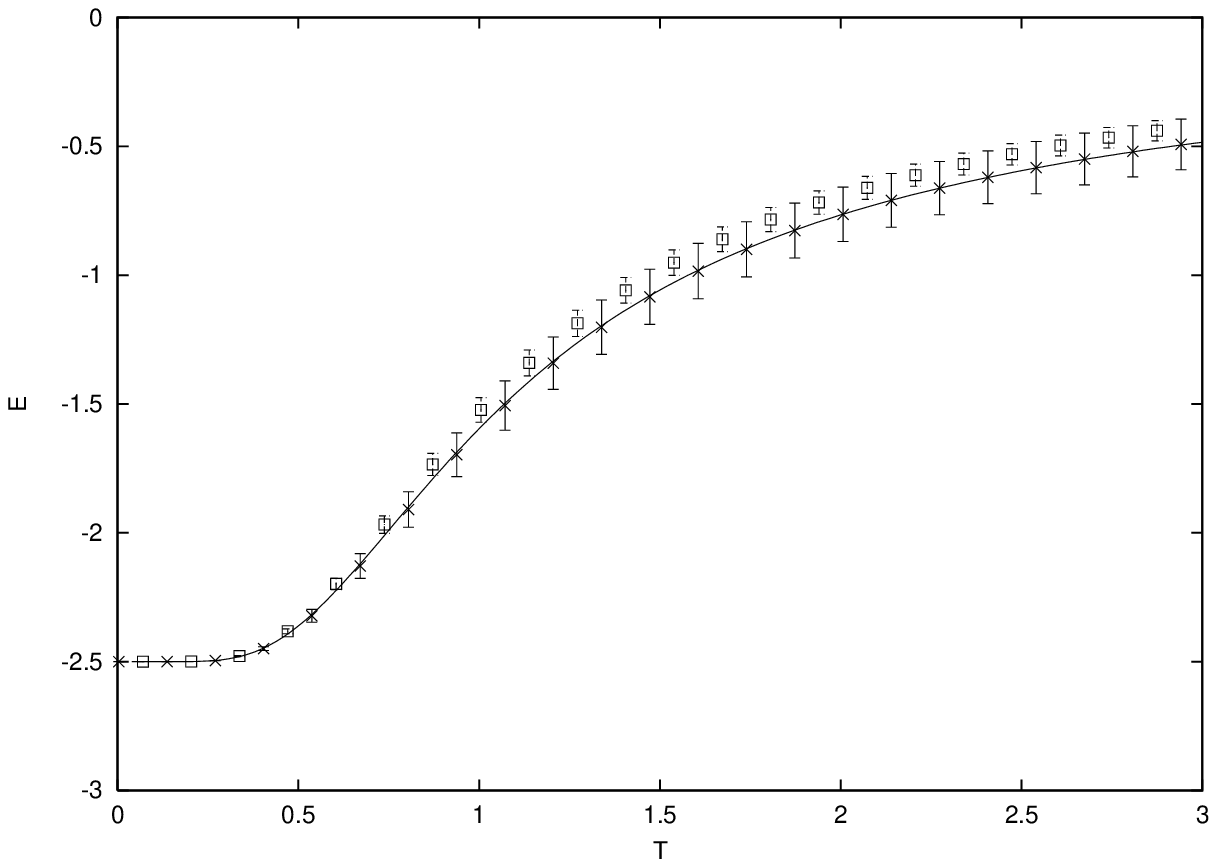}
&
\epsfxsize=5.3cm
\epsffile{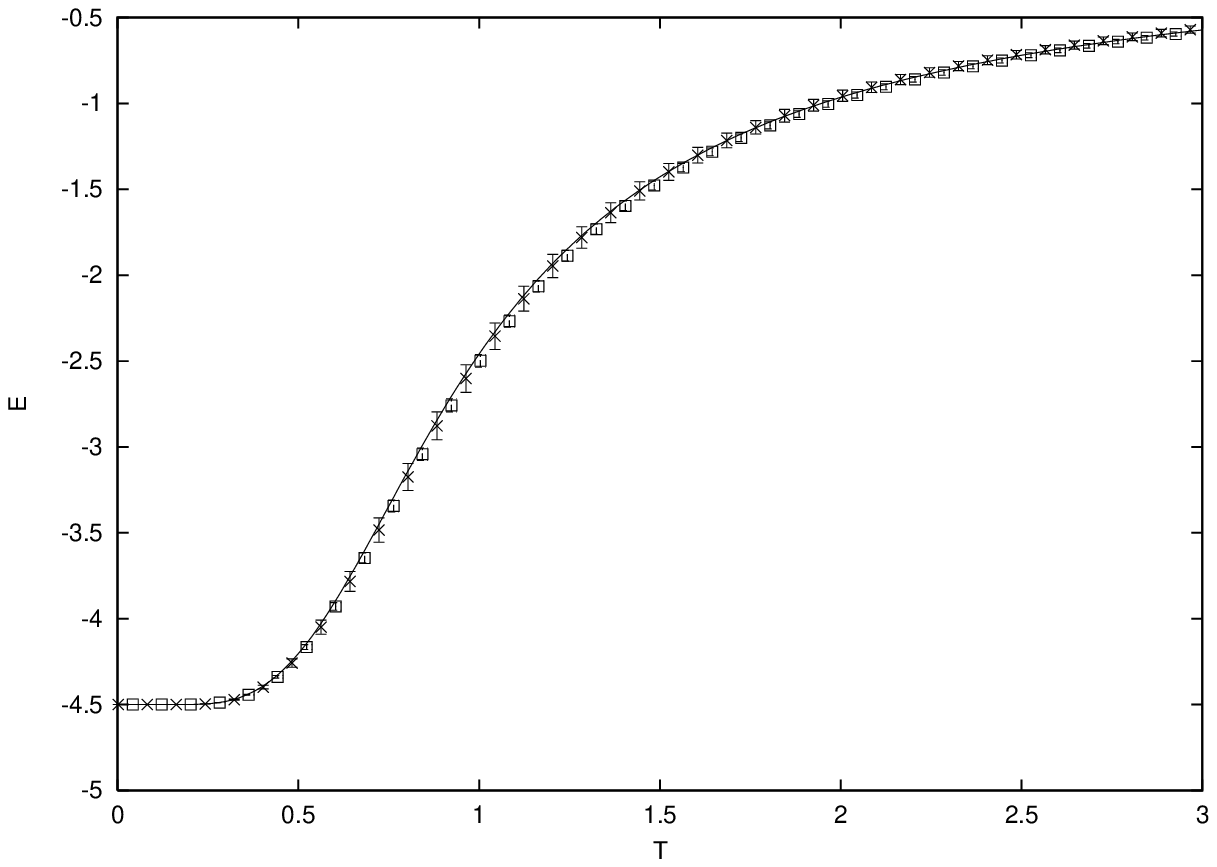}
&
\epsfxsize=5.3cm
\epsffile{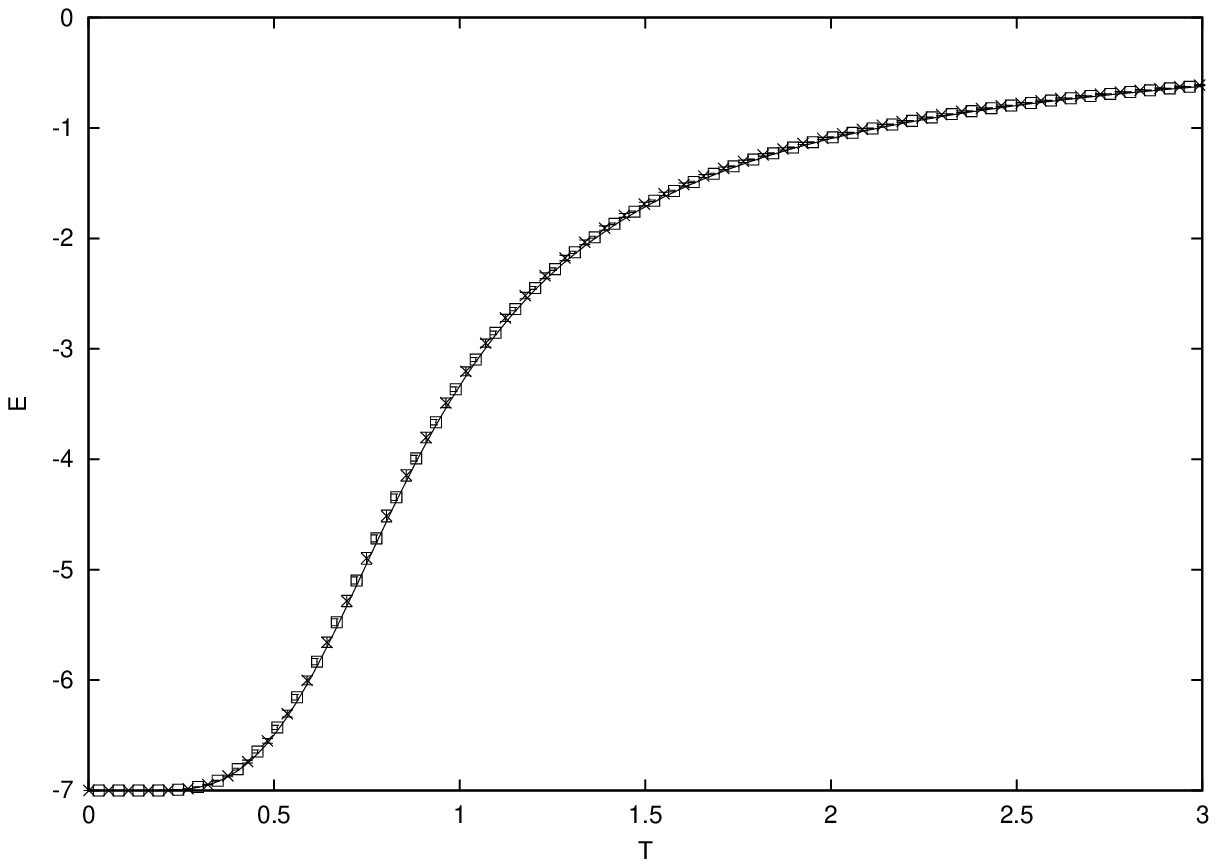}
\\
\epsfxsize=5.3cm
\epsffile{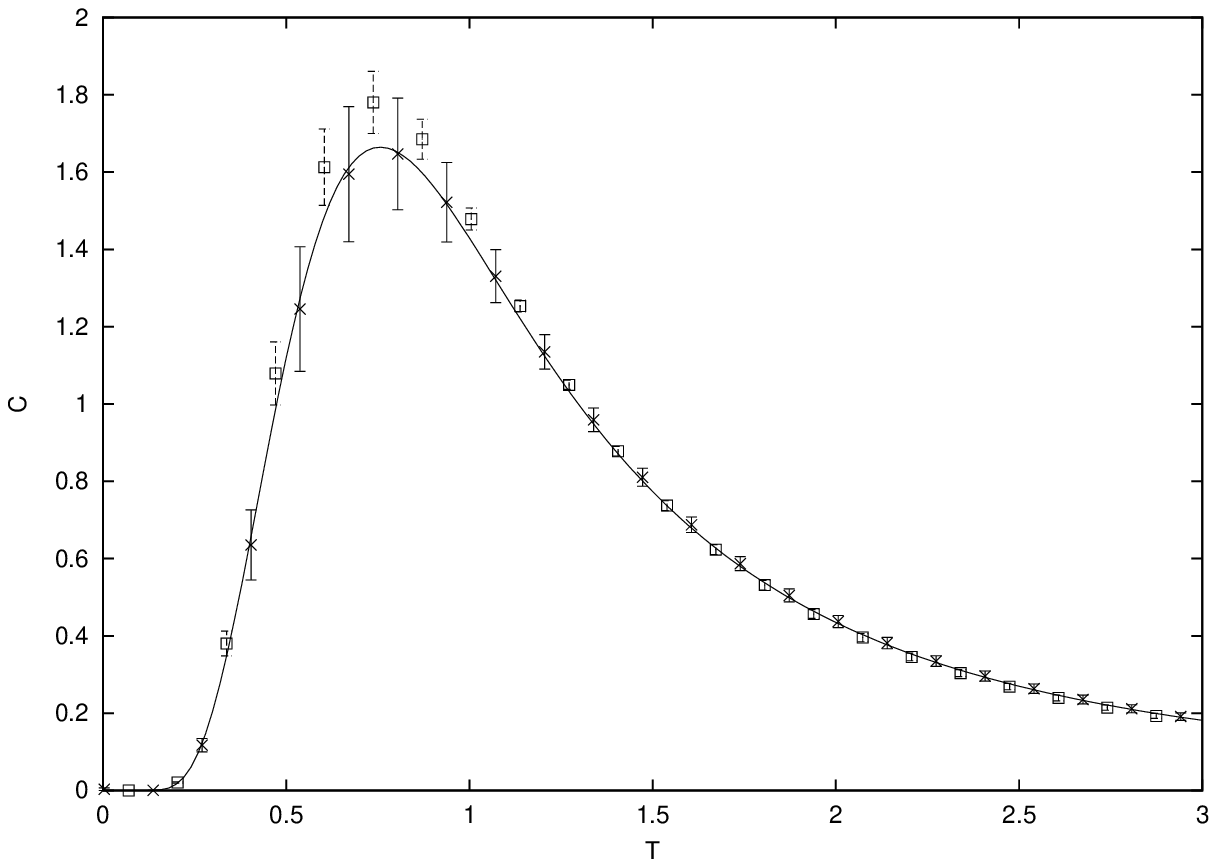}
&
\epsfxsize=5.3cm
\epsffile{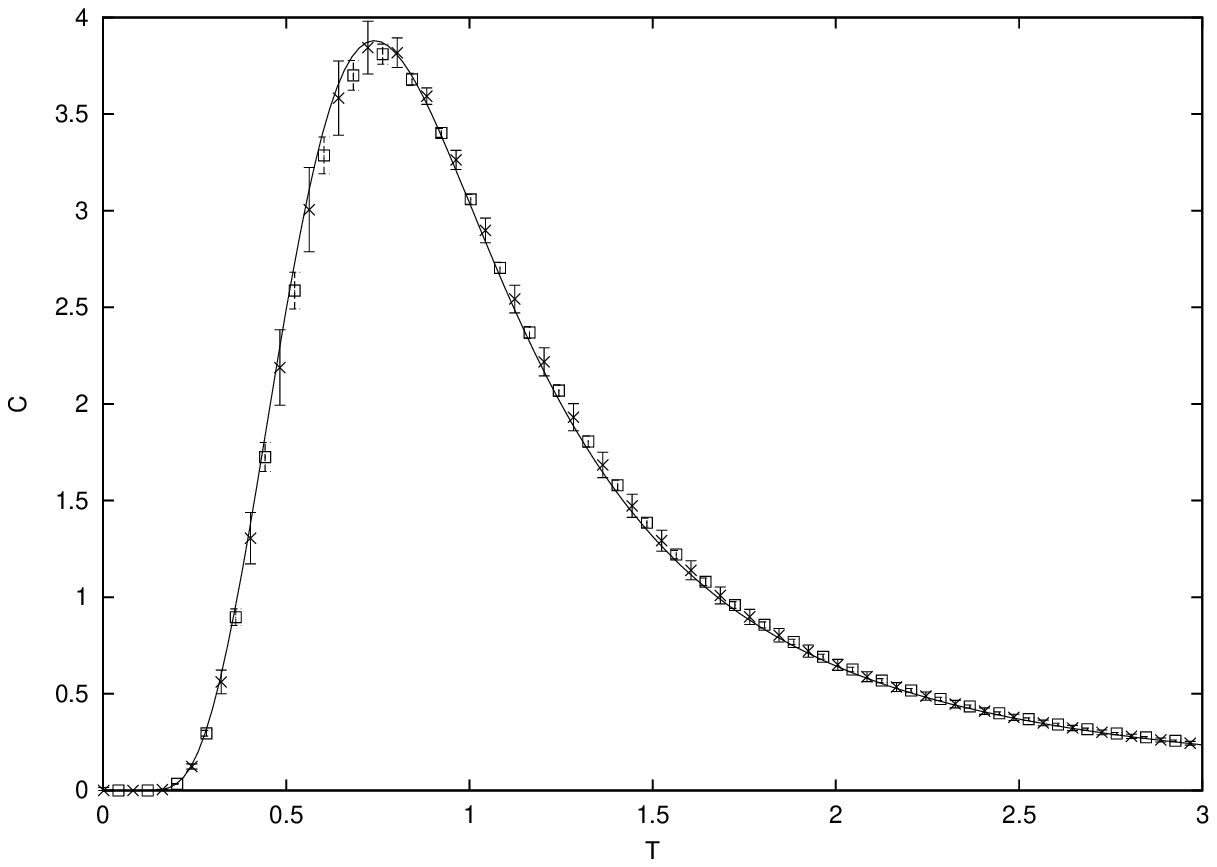}
&
\epsfxsize=5.3cm
\epsffile{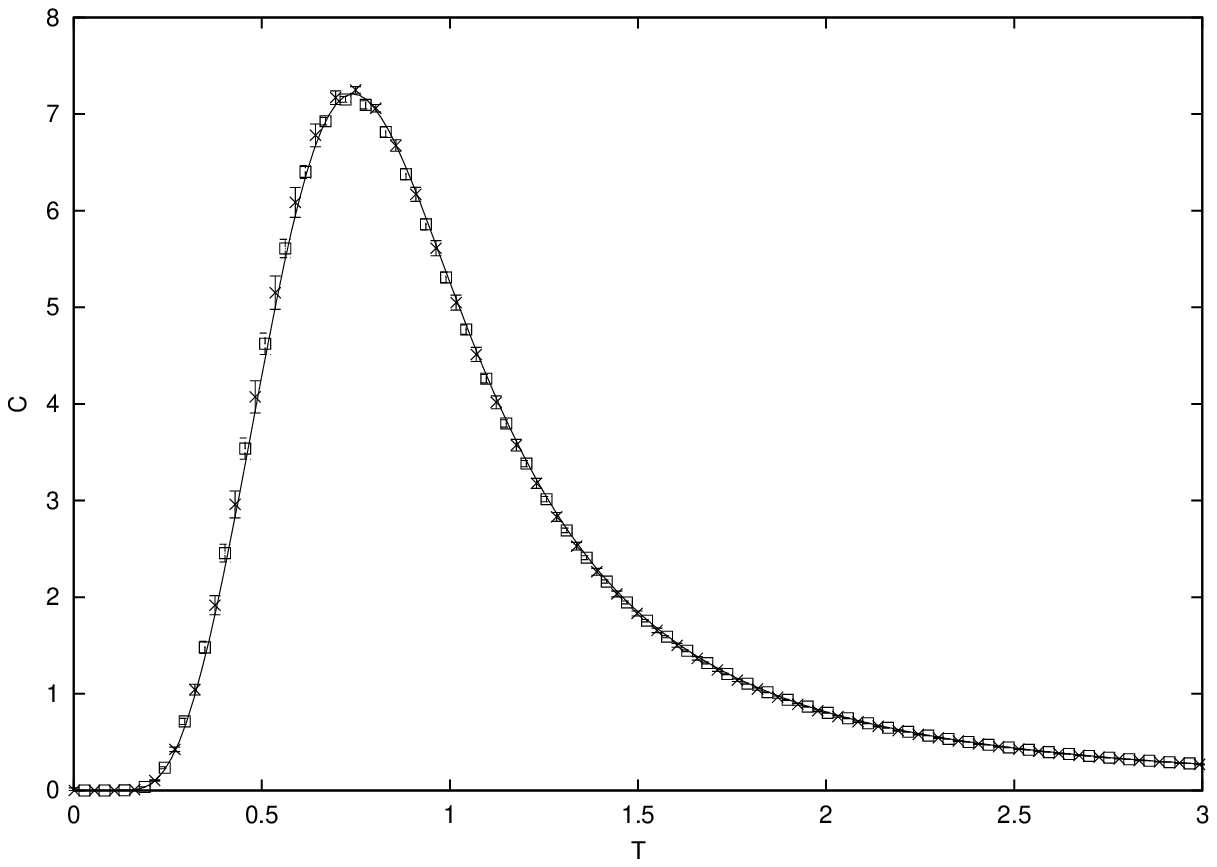}
\end{tabular}
\end{center}
\caption{
Energy (top) and specific heat (bottom)
of the mean-field model (see~(\ref{eq:mfhamil})) with $J=1$ and $h=0$.
Left: $L=6$; middle: $L=10$; right: $L=15$.
Solid lines: Exact result.
crosses: Simulation data using $S=5$ samples;
squares: Simulation data using $S=20$ samples.
Error bars: One standard deviation.
}
\label{fig:mfreal}
\end{figure}

Results for the energy $E$ and specific heat $C$ are presented in
Fig.~\ref{fig:xyreal} (XY model), \ref{fig:isreal} (Ising model in a transverse
field), and \ref{fig:mfreal} (mean-field model).
The solid lines represent the exact result for the case shown.
Simulation data as obtained from $S=5$ and $S=20$ samples are represented by symbols,
the estimates of the statistical error by error bars.
We see that the data are in excellent agreement with the exact
results and equally important,
the estimate for the error captures the deviation from the exact result very well.
We also see that in general the error decreases with the system size.
Both the imaginary and real-time method seem to work very well,
yielding accurate results for the energy and specific heat of quantum spin
systems with modest amounts of computational effort.

\section{Conclusions} \label{sec:conclusions}
The theoretical analysis presented in this paper
gives a solid justification of the remarkable efficiency of the
real-time equation-of-motion method for computing the distribution
of all eigenvalues of very large matrices.
The real-time method can be used whenever the more conventional, Lanczos-like,
sparse-matrix techniques can be applied: Memory and CPU requirements
for each iteration (time-step) are roughly the same
(depending on the actual implementation) for both approaches.

We do not recommend using the real-time method
if one is interested in the smallest (or largest) eigenvalue only.
Then the Lanczos method is computationally more efficient
because it needs less iterations (time-steps) than the real-time approach.
However if one needs information about all eigenvalues and
direct diagonalization is not possible (because of memory/CPU-time)
there is as yet no alternative to the real-time method.
The matrices used in this example (up to $32768\times32768$) are
not representative in this respect: The real-time method
has been used to compute the distribution of eigenvalues
for matrices of dimension $16777216\times16777216$~\cite{PEDROone}.

Once the eigenvalue distribution is known the thermodynamic
quantities directly follow. However if one is interested in
the accurate determination of the temperature
dependence of thermodynamic (and static correlation functions) properties
but not in the eigenvalue distribution itself, the imaginary-time method is
by far the most efficient method to compute these quantities.
For instance the calculation of the thermodynamic
properties for $\beta J=0,\ldots,10$
of a 15-site spin-1/2 system (i.e. implicitly solving the full
$32768\times32768$ eigenvalue problem) takes 1410 seconds
per sample on a Mobile Pentium III 500 Mhz system.

Finally we remark that although we used quantum-spin models to illustrate
various aspects, there is nothing in the real or imaginary-time method
that is specific to the models used. The only requirement for these
methods to be useful in practice is that the matrix is sparse and
(very) large.

\section*{Acknowledgments}
Support from the Dutch ``Stichting Nationale Computer
Faciliteiten (NCF)'' and the Dutch ``Stichting voor Fundamenteel Onderzoek der Materie (FOM)''
is gratefully acknowledged.

\appendix
\section{Expectation value calculation} \label{app:expect}
In this appendix we calculate the expectation value of the error squared,
as defined in Section~\ref{sec:theory}.
By definition we have
\widetext
 \begin{align}
E\left(\left|\Tr RA\right|^2\right) 
&=
E\left(
\left|\frac{1}{S} \sum_{p = 1}^S \sum_{m,n=1}^D \left(\delta_{m,n}-D\,c^*_{m,p} c^{\phantom{*}}_{n,p}\right) A_{m,n}\right|^2
\right)
\nonumber \\
&= 
\frac{1}{ S^2} \sum_{p,p' = 1}^S \sum_{k,l,m,n=1}^D 
\left(\delta_{k,l}\delta_{m,n} - D\,\delta_{k,l}\,E(c^*_{m,p} c^{\phantom{*}}_{n,p})
\right. \nonumber \\ & \left. -D\,\delta_{m,n}\,E(c^{\phantom{*}}_{k,p'} c^*_{l,p'})
+ D^2\,E(c^*_{m,p}\,c^{\phantom{*}}_{n,p}\,c^{\phantom{*}}_{k,p'}\,c^*_{l,p'})
\right)\,A^*_{k,l}  A_{m,n},
\end{align}
where $p$ and $p'$ label the realization of the random numbers
$c_{n,p}\equiv f_{n,p}+ i g_{n,p}$.

First we assume that different realizations $p \neq p'$ are independent
implying that
\begin{align}
E(c^*_{m,p}\,c_{n,p}\,c_{k,p'}\,c^*_{l,p'})_{p \neq p'} = 
E(c^*_{m,p}\,c_{n,p})E(c_{k,p'}\,c^*_{l,p'}).
\end{align}
Second we assume that the random numbers are
drawn from a probability distribution that
is an even function of each variable
\begin{align} \label{eq:cnreq1}
& P(f_{1,p},g_{1,p},f_{2,p},g_{2,p},\ldots,f_{k,p},g_{k,p},\ldots,f_{D,p},g_{D,p})  \nonumber \\
&=
P(f_{1,p},g_{1,p},f_{2,p},g_{2,p},\ldots,-f_{k,p},g_{k,p},\ldots,f_{D,p},g_{D,p})
\nonumber \\
&= 
P(f_{1,p},g_{1,p},f_{2,p},g_{2,p},\ldots,f_{k,p},-g_{k,p},\ldots,f_{D,p},g_{D,p}) ,
\end{align}
and symmetric under interchange of any two variables
\begin{align} \label{eq:cnreq2}
& P(f_{1,p},g_{1,p},\ldots,f_{i,p},g_{i,p},\ldots,f_{j,p},g_{j,p},\ldots,f_{D,p},g_{D,p})  \nonumber \\
&=
P(f_{1,p},g_{1,p},\ldots,f_{j,p},g_{i,p},\ldots,f_{i,p},g_{j,p},\ldots,f_{D,p},g_{D,p})
\nonumber \\
&= 
P(f_{1,p},g_{1,p},\ldots,g_{i,p},f_{i,p},\ldots,f_{j,p},g_{j,p},\ldots,f_{D,p},g_{D,p}),
\end{align}
for all $i,j,k=1,\ldots,D$.
This is most easily done by drawing individual numbers
from the same even probability distribution i.e.
\begin{align} \label{eq:cnreq3}
P(f_{1,p},g_{1,p},\ldots,f_{j,p},g_{i,p},\ldots,f_{i,p},g_{j,p},\ldots,f_{D,p},g_{D,p})
&=
\prod_{n,m=1}^{D} P(f_{n,p})P(g_{n,p}),
\end{align}
where $P(x)=P(-x)$.
Normalizing the vector $(f_{1,p},g_{1,p},\ldots,f_{D,p},g_{D,p})$
such that $\sum_{i=1}^D |c_{n,p}|^2 =1$ (for $p=1,\ldots,S$)  does not affect
the basic requirements~(\ref{eq:cnreq1}) and~(\ref{eq:cnreq2}).

Making use of the above properties of $P(f_1,g_1,\ldots,f_D,g_D)$
we find that
\begin{align}
E(c^*_{m,p} c^{\phantom{*}}_{n,p}) = \delta_{m,n} E(|c_{m,p}|^2)=\delta_{m,n} E(|c|^2),
\end{align}
where in the last equality we omitted the subscripts of $c_{m,p}$
to indicate that the expectation value does not depend on $m$ or $p$.
An expectation value of a product of two $c^*$'s and two $c$'s can be
written as
\widetext
\begin{align} \label{eq:fourpoint}
E(c^*_{m,p}\,c^{\phantom{*}}_{n,p}\,c^{\phantom{*}}_{k,p'}\,c^*_{l,p'})
=&
(1-\delta_{p,p'})\delta_{m,n}\delta_{k,l} E(|c_{m,p}|^2) E(|c_{m,p'}|^2)
\nonumber \\ &
+
\delta_{p,p'} \delta_{m,n}\delta_{k,l} (1-\delta_{mk})
E(c^*_{m}\,c^{\phantom{*}}_{m}\,c^{\phantom{*}}_{k}\,c^*_{k})
\nonumber \\ &
+\delta_{p,p'} \delta_{m,k}\delta_{n,l} (1-\delta_{m,n})
E(c^*_{m}\,c^{\phantom{*}}_{n}\,c^{\phantom{*}}_{m}\,c^*_{n})
\nonumber \\ &
+\delta_{p,p'} \delta_{m,l}\delta_{n,k} (1-\delta_{m,n})
 E(c^*_{m}\,c^{\phantom{*}}_{n}\,c^{\phantom{*}}_{n}\,c^*_{m})
\nonumber \\ &
+\delta_{p,p'} \delta_{m,l}\delta_{n,k}\delta_{m,n}
E(c^*_{m}\,c^{\phantom{*}}_{m}\,c^{\phantom{*}}_{m}\,c^*_{m})
\nonumber \\ 
=&
(1-\delta_{p,p'})\delta_{m,n}\delta_{k,l} E(|c|^2)^2
\nonumber \\ &
+\delta_{p,p'} \delta_{m,n}\delta_{k,l} (1-\delta_{m,k})
E(|c_{m,p}|^2\,|c_{k,p}|^2)
\nonumber \\ &
+\delta_{p,p'} \delta_{m,k}\delta_{n,l} (1-\delta_{m,n})
 E(|c_{m,p}|^2\,|c_{n,p}|^2)
\nonumber \\ &
+\delta_{p,p'} \delta_{m,l}\delta_{n,k} (1-\delta_{m,n})
E(c^*_{m,p}\,c^{\phantom{*}}_{n,p}\,c^{\phantom{*}}_{n,p}\,c^*_{m,p})
\nonumber \\ &
+\delta_{p,p'} \delta_{m,l}\delta_{n,k}\delta_{m,n} E(|c_{m,p}|^4).
\end{align}
Furthermore for $m\neq n$ we have
\begin{align}
E(c^*_{m,p}\,c^{\phantom{*}}_{n,p}\,c^{\phantom{*}}_{n,p}\,c^*_{m,p})=&
E(
(f^2_{m,p} - 2 i f_{m,p} g_{m,p} - g_{m,p}^2)
(f^2_{n,p} + 2 i f_{n,p} g_{n,p} - g_{n,p}^2)
)
\nonumber \\
=&
 E(f^2_{m,p} f^2_{n,p}) + 2 i E(f^2_{m,p} f_{n,p} g_{n,p}) - E(f^2_{m,p} g_{n,p}^2)
\nonumber \\
&
- 2 i E(f_{m,p} g_{m,p} f^2_{n,p})
+ 4 E(f_{m,p} f_{n,p} g_{m,p}  g_{n,p}) + 2 i E(f_{m,p} g_{m,p} g_{n,p}^2)
\nonumber \\
&
 -E(g_{m,p}^2 f^2_{n,p}) - 2 i E(g_{m,p}^2 f_{n,p} g_{n,p}) + E(g_{m,p}^2 g_{n,p}^2)
\nonumber \\
=&
E(f^2_{m,p} f^2_{n,p}) 
- E(g_{m,p}^2 f^2_{n,p})-E(f^2_{m,p} g_{n,p}^2)+ E(g_{m,p}^2  g_{n,p}^2) 
\nonumber \\
=&
 0.
\end{align}
By symmetry $ E(|c_{m,p}|^2\,|c_{n,p}|^2)$
does not depend on $m$, $n$ or $p$ and the same holds for $E(|c_{m,p}|^4)$.

The fact that the vector
$(c_{1,p},\ldots,c_{D,p})$ is normalized yields the identities
\begin{align}
E\left(\sum_{n=1}^D |c_{n,p}|^2 \right) =
\sum_{n=1}^D E(|c_{n,p}|^2) = D E(|c|^2) = E(1) = 1 ,
\end{align}
and
\begin{align}
E\left( \left(\sum_{n=1}^D |c_{n,p}|^2 \right)^2\right)
&= \sum_{m,n=1}^D E(|c_{n,p}|^2 |c_{m,p}|^2)
\nonumber \\
&= \sum_{n=1}^D E(|c_{n,p}|^4)
+\sum_{m,n=1}^D (1-\delta_{m,n}) E(|c_n|^2 |c_m|^2)
\nonumber  \\
&=
D E(|c|^4) + D(D-1) E(|c|^2 |c^\prime|^2)= E(1) =1 ,
\end{align}
where $c$ and $c^\prime$ refer to two different complex random variables.
Therefore we have
\begin{align}
 E(|c|^2) &= 1/D, 
\end{align}
and
\begin{align}
 E(|c|^2 |c^\prime|^2) &= \frac{1 - D E(|c|^4)}{D (D-1)}.
\end{align}
Substitution into~(\ref{eq:fourpoint}) yields
\begin{align}
E(c^*_{m,p}\,c_{n,p}\,c_{k,p'}\,c^*_{l,p'}) 
=&
(1-\delta_{p,p'})\delta_{m,n}\delta_{k,l} D^{-2}
\nonumber \\
&+
\delta_{p,p'} \frac{1 - D E(|c|^4)}{D (D-1)}
\left(\delta_{m,n}\delta_{k,l} (1-\delta_{m,k})
+ \delta_{m,k}\delta_{n,l} (1-\delta_{m,n}) \right)
\nonumber 
\\
&+\delta_{p,p'} \delta_{m,l}\delta_{n,k}\delta_{m,n} E(|c|^4).
\end{align}
and the final result for the variance reads
\begin{align} \label{eq:generalerror}
E\left(\left|\Tr RA\right|^2\right) 
=&
\frac{1}{ S}  
\left(
\frac{D - D^2 E(|c|^4)}{D-1}  \Tr A^\dagger A 
+ \frac{1 - D^2 E(|c|^4)}{D-1}   |\Tr A|^2 
\right. \nonumber \\ & \left. + \frac{(D+1) D^2 E(|c|^4) - 2 D }{D-1}  \sum_{n=1}^D |A_{n,n}|^2
\right).
\end{align}

An expression for the fourth moment $E(|c|^4)$ cannot be derived from general
properties of the probability distribution or normalization of random vector.
We can only make progress by specifying the former explicitly.
As an example we take a probability distribution such that
for each realization $p$ the random numbers $f_{n,p}$ and $g_{n,p}$ are
distributed uniformly over the surface of a $2D$-dimensional sphere of radius $1$.
This probability distribution can be written as
\begin{align}
P(f_{1},g_{1},f_{2},g_{2},\ldots,f_{D},g_{D})
\propto \delta(f_{1}^2+g_{1}^2+f_{2}^2+g_{2}^2+\ldots +f_{D}^2+g_{D}^2 - 1),
\end{align}
where we omitted the subscript $p$ because it is
irrelevant for what follows.
The even moments of $|c_{n}|=(f_{n}^2+g_{n}^2)^{1/2}$ are defined by
\begin{align}
E(|c|^{2M}) =
\frac{ \int_{-\infty}^{\infty} 
(f_{1}^2+g_{1}^2)^{M} \delta(f_{1}^2+g_{1}^2+f_{2}^2+g_{2}^2+\ldots +f_{D}^2+g_{D}^2 - 1) df_{1}dg_{1}df_{2}dg_{2}\ldots df_{D}dg_{D}}
{\int_{-\infty}^{\infty} \delta(f_{1}^2+g_{1}^2+\ldots +f_{D}^2+g_{D}^2 - 1) df_{1}dg_{1}\ldots df_{D}dg_{D}}.
\end{align}
It is expedient to introduce an auxiliary integration variable $X$ by
\begin{align} \label{eq:xmoment}
E(|c|^{2M}) =
\frac{ \int_{-\infty}^{\infty} 
X^{M} \delta(f_{1}^2+g_{1}^2-X)\delta(f_{2}^2+g_{2}^2+\ldots +f_{D}^2+g_{D}^2 - (1-X)) dX df_{1}dg_{1}df_{2}dg_{2}\ldots df_{D}dg_{D}}
{\int_{-\infty}^{\infty} \delta(f_{1}^2+g_{1}^2+\ldots +f_{D}^2+g_{D}^2 - 1) df_{1}dg_{1}\ldots df_{D}dg_{D}}.
\end{align}
We can perform the integration over $X$ last and regard~(\ref{eq:xmoment}) as
the $M$-th moment of the variable $X$ with respect to the probability
distribution
\begin{align}
P(X) = 
 \frac{ \int_{-\infty}^{\infty} \delta(f_{1}^2+g_{1}^2-X) 
\delta(f_{2}^2+g_{2}^2+\ldots +f_{D}^2+g_{D}^2 -( 1-X)) df_{1}dg_{1}df_{2}dg_{2}\ldots df_{D}dg_{D}}
{\int_{-\infty}^{\infty} \delta(f_{1}^2+g_{1}^2+\ldots +f_{D}^2+g_{D}^2 - 1) df_{1}dg_{1}\ldots df_{D}dg_{D}}.
\end{align}
The calculation of $P(X)$ amounts to computing integrals of the form
\begin{align}
I_N(X) = \int_{-\infty}^{\infty} \delta\left(\sum_{n=1}^N x_n^2 -X\right) dx_1dx_2\ldots dx_N.
\end{align}
Changing to spherical coordinates we have
\begin{align}
I_N(X) 
&= \frac{2 \pi^{N/2}}{\Gamma(N/2)} \int_{0}^{\infty} r^{N-1} \delta(r^2-X) dr 
\nonumber \\
&= \frac{\pi^{N/2}}{\Gamma(N/2)} X^{N/2-1} \theta(X),
\end{align}
yielding
\begin{align}
P(X) &= \frac{I_{2}(X) I_{2D-2}(1-X)}{I_{2D}(1)}
\nonumber \\ &
= (D-1) (1-X)^{D-2} \theta(X) \, \theta (1-X).
\end{align}
The moments $E(|c|^{2M})$ are given by
\begin{align}
E(|c|^{2M}) &= \int_{-\infty}^{\infty} X^{M} P(X) d X
\nonumber \\ &
=  (D-1) \int_0^1  X^{M} (1-X)^{D-2} dX
\nonumber \\ &
= 
\frac{\Gamma(D)\Gamma(1+M)}{\Gamma(D+M)},
\end{align}
and the values of interest to us are
\begin{align} \label{eq:isomoments}
E(|c|^0) = 1, \quad
E(|c|^2) = \frac{1}{D}, \quad
E(|c|^4) = \frac{2}{D (D+1)}, 
\end{align}
where the first two results provide some check on the procedure used.
Substituting~(\ref{eq:isomoments}) into~(\ref{eq:generalerror}) yields
\begin{align}
E\left(\left|\Tr RA\right|^2\right) 
&= 
\frac{D \Tr A^\dagger A -  |\Tr A|^2 }{ S (D+1)}  
.
\end{align}

\section{Error Bounds} \label{app:bounds}

Here we prove that the difference between the eigenvalues of the
Hermitian matrix $A+B$
and those obtained from the approximate time-evolution
$\exp(z A/2)\exp(z B)\exp(z A/2)$ ($z=-i\tau,-\tau$)
is bounded by $\tau^2$. In the following we assume $A$ and $B$ are Hermitian
matrices and take $\tau$ a real, non-negative number.
We start with the imaginary-time case.

We define the difference $R(\tau)$ by
\begin{align}
R(\tau) \equiv& e^{\tau(A+B)} -e^{\tau A/2}e^{\tau B}e^{\tau A/2}
\nonumber \\
=&
 \frac{1}{4}
\int_0^\tau d \lambda
\int_0^\lambda d \mu
\int_0^\mu d \nu
e^{\lambda A/2}
e^{\lambda B}
\{ 
e^{-\nu B} [2 B,[A,B]] e^{\nu B} 
\nonumber  \\
& +e^{\nu A/2} [A,[A,B]] e^{-\nu A/2}
\}
e^{\lambda A/2}
e^{(\tau-\lambda) (A+B)},
\end{align}
a well-known result \cite{SUZUKItwo2}.
We have~\cite{SUZUKIthree}
\begin{align}
|| R(\tau) || 
\leq&
\frac{1}{4} \left|\left|
\int_0^\tau d \lambda
\int_0^\lambda d \mu
\int_0^\mu d \nu
e^{\lambda A/2}
e^{(\lambda-\nu) B}  [2 B,[A,B]] e^{\nu B} e^{\lambda A/2} e^{(\tau-\lambda) (A+B)}
\right|\right|
\nonumber \\
&+ \frac{1}{4} \left|\left|
\int_0^\tau d \lambda
\int_0^\lambda d \mu
\int_0^\mu d \nu
e^{\lambda A/2}
e^{\lambda B}
e^{\nu A/2} [A,[A,B]] e^{(\lambda-\nu) A/2} e^{(\tau-\lambda) (A+B)}
\right|\right|
\nonumber \\
\leq &
\frac{1}{4} 
\int_0^{\tau} d \lambda
\int_0^\lambda d \mu
\int_0^\mu d \nu
e^{\lambda ||A||/2}
e^{(\lambda-\nu) ||B||} || [2 B,[A,B]] ||e^{\nu ||B||} e^{\lambda ||A||/2} e^{(\tau-\lambda) (||A||+||B||)}
\nonumber \\
&+ \frac{1}{4}
\int_0^{\tau} d \lambda
\int_0^\lambda d \mu
\int_0^\mu d \nu
e^{\lambda ||A||/2}
e^{\lambda ||B||}
e^{\nu ||A||/2} ||[A,[A,B]]|| e^{(\lambda-\nu) ||A||/2} e^{(\tau-\lambda) (||A||+||B||)}
\nonumber \\
&=  
\frac{1}{24} \tau^3 e^{\tau (||A||+||B||)}
\left(||[A,[A,B]]|| 
+|| [2 B,[A,B]] ||\right),
\end{align}
and
\begin{align} 
|| R(-\tau) || 
\leq &
\frac{1}{4} \left|\left|
\int_0^{-\tau} d \lambda
\int_0^\lambda d \mu
\int_0^\mu d \nu
e^{\lambda A/2}
e^{(\lambda-\nu) B}  [2 B,[A,B]] e^{\nu B} e^{\lambda A/2} e^{(-\tau-\lambda) (A+B)}
\right|\right|
\nonumber \\
&+ \frac{1}{4} \left|\left|
\int_0^{-\tau} d \lambda
\int_0^\lambda d \mu
\int_0^\mu d \nu
e^{\lambda A/2}
e^{\lambda B}
e^{\nu A/2} [A,[A,B]] e^{(\lambda-\nu) A/2} e^{(-\tau-\lambda) (A+B)}
\right|\right|
\nonumber \\
=&
\frac{1}{4} \left|\left|
\int_0^\tau d \lambda
\int_0^{\lambda} d \mu
\int_0^{\mu} d \nu
e^{-\lambda A/2}
e^{(-\lambda+\nu) B}  [2 B,[A,B]] e^{-\nu B} e^{-\lambda A/2} e^{(-\tau+\lambda) (A+B)}
\right|\right|
\nonumber \\
&+ \frac{1}{4} \left|\left|
\int_0^\tau d \lambda
\int_0^{\lambda} d \mu
\int_0^{\mu} d \nu
e^{-\lambda A/2}
e^{-\lambda B}
e^{-\nu A/2} [A,[A,B]] e^{(-\lambda+\nu) A/2} e^{(-\tau+\lambda) (A+B)}
\right|\right|
\nonumber \\
\leq &
\frac{1}{4} 
\int_0^{\tau} d \lambda
\int_0^\lambda d \mu
\int_0^\mu d \nu
e^{\lambda ||A||/2}
e^{(\lambda-\nu) ||B||} || [2 B,[A,B]] ||e^{\nu ||B||} e^{\lambda ||A||/2} e^{(\tau-\lambda) (||A||+||B||)}
\nonumber \\
&+ \frac{1}{4}
\int_0^{\tau} d \lambda
\int_0^\lambda d \mu
\int_0^\mu d \nu
e^{\lambda ||A||/2}
e^{\lambda ||B||}
e^{\nu ||A||/2} ||[A,[A,B]]|| e^{(\lambda-\nu) ||A||/2} e^{(\tau-\lambda) (||A||+||B||)}
\nonumber \\
=&
\frac{1}{24} \tau^3 e^{\tau (||A||+||B||)}
\left(||[A,[A,B]]|| 
+|| [2 B,[A,B]] ||\right).
\end{align}
Hence the bound in $R(\tau)$ does not depend on the sign of $\tau$
so that we can write
\begin{align}  \label{eq:rnorm}
|| R(\tau) || 
\leq &
s |\tau|^3 e^{|\tau| (||A||+||B||)},
\end{align}
where
\begin{align}  
s \equiv \frac{1}{24} ||[A,[A,B]]|| +|| [2 B,[A,B]] ||.
\end{align}
For real $\tau$ we have
\begin{align} \label{eq:approx1}
e^{\tau A/2}e^{\tau B}e^{\tau A/2} \equiv e^{\tau C(\tau)},
\end{align}
where $C(\tau)$ is Hermitian. Clearly we have
\begin{align} \label{eq:approx2}
e^{\tau (A+B)} - e^{\tau C(\tau)} = R(\tau).
\end{align}
We already have an upperbound on $R(\tau)$ and now want to use this knowledge
to put an upperbound on the difference in eigenvalues of $C(\tau)$ and $A+B$.
In general, for two Hermitian matrices $U$ and $V$
with eigenvalues $\{u_n\}$ and $\{v_n\}$
respectively, both sets sorted in non-decreasing order, we have~\cite{WILKINSON}
\begin{align} \label{eq:eigendiff}
|u_n-v_n| \leq ||U-V|| , \quad \forall\,n.
\end{align}
Denoting the eigenvalues of $A+B$ and $C(\tau)$ by $x_n(0)$ and $x_n(\tau)$ respectively,
combining Eq.~(\ref{eq:rnorm}) and~(\ref{eq:eigendiff}) yields
\begin{align}
|e^{\tau x_n(0)} - e^{\tau x_n(\tau)}| \leq  s |\tau|^3 e^{|\tau| (||A||+||B||)}
.
\end{align}
To find an upperbound on $|x_n(0) -x_n(\tau)|$ we first assume
that $x_n(0) \leq x_n(\tau)$ and take $\tau\geq0$. It follows from (B9) that
\begin{align}
e^{\tau (x_n(\tau)-x_n(0))}-1 \leq  s \tau^3 e^{ \tau (||A||+||B||) -\tau x_n(0)} 
,
\end{align}
For $x\geq0$, $e^x-1 \geq x$ and we have
$-x_n(0) \leq  ||A+B|| \leq ||A|| +||B||$. Hence we
find
\begin{align}
x_n(\tau)-x_n(0) \leq  s \tau^2 e^{2 \tau (||A||+||B||)} 
.
\end{align}
An upperbound on the difference in the eigenvalues between $C(\tau)$ and $A+B$ can
equally well be derived by considering
the inverse of the exact and approximate time-evolution operator ~(\ref{eq:approx1}).
This is useful for the case $x_n(0) > x_n(\tau)$: Instead of using
(B7) we start from $\exp(-\tau(A+B))-\exp(-\tau C(-\tau))=R(-\tau)$ ($\tau\geq0$).
Note that the set of eigenvalues of a matrix and its inverse are the same
and that the matrices we are considering here, i.e. matrix exponentials, are nonsingular.
Making use of Eq.~(\ref{eq:rnorm}) for $R(-\tau)$ gives
\begin{align}
|e^{-\tau x_n(0)} - e^{-\tau x_n(\tau)}| \leq  s |\tau|^3 e^{|\tau| (||A||+||B||)}
,
\end{align}
and proceeding as before we find
\begin{align}
\tau (x_n(0)-x_n(\tau)) &
\leq e^{\tau (x_n(0)-x_n(\tau))}-1 \leq  s \tau^3 e^{2 \tau (||A||+||B||)}.
\end{align}
Putting the two cases together we finally have
\begin{align}  \label{eq:imagbound}
|x_n(\tau)-x_n(0)| \leq  s \tau^2 e^{2 \tau (||A||+||B||)} 
.
\end{align}
Clearly (B14) proves that the differences in the eigenvalues
of $A+B$ and $C(\tau)$ vanish as $\tau^2$.

We now consider the case of the real-time algorithm ($z=-i\tau$).
For Hermitian matrices $A$ and $B$ the matrix exponentials are unitary matrices
and hence their norm equals one. This simplifies the
derivation of the upperbound on $R(-i\tau)$.
One finds~\cite{HDRCPR}
\begin{align}  \label{eq:rnorm2}
|| R(-i\tau) ||_E
&\leq 
s |\tau|^3 
,
\end{align}
where $||A||^2_E \equiv \Tr A^\dagger A$ denotes the Euclidean norm
of the matrix $A$~\cite{WILKINSON}.
In general the eigenvalues of a unitary matrix are complex valued
and therefore the strategy adopted above to use the bound on $R(\tau)$ to
set a bound on the difference of the eigenvalues no longer works.
Instead we invoke the Wielandt-Hoffman theorem~\cite{Hoffman}:

\noindent
{\sl If $U$ and $V$ are normal matrices
with eigenvalues $u_i$ and $v_i$ respectively,
then there exists a suitable rearrangement
(a permutation $\perm$ of the numbers $1,\ldots,n$)  of the
eigenvalues so that
\begin{align}
\sum_{j=1}^N |u_j-v_{\perm(j)}|^2 \leq ||U-V||^2_E\, .
\end{align}
}

\noindent
Let $U$ and $V$ denote the exact and approximate real-time evolution operators
respectively.
The eigenvalues of $A+B$ and $ C(\tau)$ are $x_n(0)$ and $x_n(\tau)$ respectively.
All the $x_n$'s and $x_n(\tau)$'s are real numbers.
According to the Wielandt-Hoffman theorem
\begin{align} \label{eq:ineq1}
\sum_{j=1}^N |e^{i\,\tau\,x_j(0)}-e^{i\,\tau\,y_j(\tau)}|^2 \leq ||R(-i\tau)||^2_E \leq s^2 \tau^6 .
\end{align}
where $y_j(\tau)=x_{\perm(j)}(\tau)$, $\perm$ being the permutation such that
inequality~(\ref{eq:ineq1}) is satisfied.
We see that Eq.~(\ref{eq:ineq1}) only depends on $(\tau\,x_j(0) \mod 2 \pi)$ and $(\tau\,y_j(\tau) \mod 2 \pi)$,
but so does the DOS (see Eq.~(\ref{eq:dos})). 
Since the inequality (B17) and the DOS
only depend on these ``angles'' modulo $2\pi$,
there is no loss of generality if we make the choice
\begin{align} \label{eq:diffrange}
0\leq | \tau(x_j(0)-y_j(\tau))| \leq \pi.
\end{align}
Rewriting the sum in (B17) we have
\begin{align}
\sum_{j=1}^N |e^{i\,\tau\,x_j(0)}-e^{i\,\tau\,y_j(\tau)}|^2 
&= \sum_{j=1}^N 
(
2-2\,\cos(\tau\,(x_j(0)-y_j(\tau)))
) \nonumber \\ 
&= 4 \sum_{j=1}^N 
\sin^2(\tau/2\,(x_j(0)-y_j(\tau))).
\end{align}
As we have
\begin{align}
\sin^2 x \leq \frac{4\,x^2}{\pi^2} , \text{ for } 0 \leq |x| \leq \pi/2,
\end{align}
the restriction Eq.~(\ref{eq:diffrange}) allows us to write
\begin{align}
 \sum_{j=1}^N (x_j(0)-y_j(\tau))^2 \leq \frac{\pi^2 s^2}{4} \tau^4,
\end{align}
implying
\begin{align} \label{eq:realbound}
|x_j(0)-y_j(\tau)| \leq \frac{\pi s}{2} \tau^2.
\end{align}
In summary, we have shown that in the real-time case there exists a
permutation of the approximate eigenvalues
such that the difference with the exact ones vanishes as $\tau^2$.

Finally we note that both upperbounds~(\ref{eq:realbound}) and~(\ref{eq:imagbound})
hold for arbitrary Hermitian matrices $A$ and $B$ and are therefore rather weak.
Except for the fact that they provide a sound theoretical justification
for the real and imaginary-time method,
they are of little practical value.

%\begin{figure} 
%\epsfxsize=12.0cm
%\begin{center}\epsffile{fig2.eps}\end{center}
%\caption{ Energy per site as a function of the temperature for
% triangular lattices of different size $L$.
% Lines represent data obtained by the quantum algorithm described in the text,
% using $S=20$ samples .
% Markers represent results obtained by exact diagonalization of the Hamiltonian.}
%\label{fig:eofl}
%\end{figure}

\onecolumn

\pagebreak
% \begin{figure}[p]
% \begin{center}
% \begin{tabular}{ccc}
% \epsfxsize=5.3cm
% \epsffile{figures/xy6h-E.eps}
% &
% \epsfxsize=5.3cm
% \epsffile{figures/is6h-E.eps}
% &
% \epsfxsize=5.3cm
% \epsffile{figures/mf6h-E.eps}
% \end{tabular}
% \end{center}
% \caption{
% Energy as a function of temperature calculated with the unbiased random generator,
% for the XY model, the Ising model in a transverse field, and the Mean field model.
% }
% \label{fig:homo}
% \end{figure}


\end{document}